\documentclass[12pt,preprint]{aastex}
\usepackage{natbib}
\usepackage{graphicx}
\usepackage{amssymb}
\usepackage{amsmath}
\shorttitle{Constraining annihilating dark matter mass}
\begin{document}
\title{Constraining annihilating dark matter mass by the radio continuum spectral data of a high-redshift galaxy cluster}
\author{Man Ho Chan$^1$, Chak Man Lee$^1$, C.-Y. Ng$^2$, Chun Sing Leung$^3$}
\affil{$^1$Department of Science and Environmental Studies, The Education University of Hong Kong, Hong Kong, China \\
$^2$ Department of Physics, The University of Hong Kong, Hong Kong, China \\
$^3$ Department of Applied Mathematics, Hong Kong Polytechnic University, Hong Kong, China}
\email{chanmh@eduhk.hk}

\begin{abstract}
In the past decade, the properties of annihilating dark matter models were examined by various kinds of data, including the data of gamma rays, radio waves, X-ray, positrons, electrons, antiprotons and neutrinos. In particular, most of the studies focus on the data of our Galaxy, nearby galaxies (e.g. M31 galaxy) or nearby galaxy clusters (e.g. Fornax cluster). In this article, we examine the archival radio continuum spectral data of a relatively high-redshift galaxy cluster (A697 cluster) to constrain the properties of annihilating dark matter. We find that leptophilic annihilation channels ($e^+e^-$, $\mu^+\mu^-$ and $\tau^+\tau^-$) can give very good fits to the radio continuum spectrum of the A697 cluster.   
\end{abstract}


\section{Introduction}
Recently, many studies have provided severe constraints on the properties of annihilating dark matter. For example, several detectors such as Fermi-LAT, AMS and DAMPE were launched to detect signals of annihilating dark matter in our Galaxy. Some excess of gamma rays \citep{Daylan,Calore}, anti-protons \citep{Cholis} and electrons/positrons \citep{Ambrosi,Aguilar} have been reported and claimed as dark matter signals. In particular, one popular dark matter interpretation of the alleged excess suggests the existence of annihilating dark matter with mass $m \sim 30-80$ GeV \citep{Daylan,Calore,Cholis}. The best-fit annihilation cross section ranges are close to the thermal annihilation cross section $\sigma v=2.2\times 10^{-26}$ cm$^3$ s$^{-1}$ predicted by standard cosmology \citep{Steigman}. Later, analyses of radio halos of galaxy clusters also support this suggestion \citep{Chan,Chan2}. Although some recent gamma-ray studies of the Omega Centauri and 47 Tuc clusters suggest a slightly smaller best-fit dark matter mass range $m \approx 30-35$ GeV and smaller annihilation cross sections \citep{Brown,Brown2}, the range $m \sim 30-80$ GeV has become one of the most attentive possible ranges of annihilating dark matter mass. Interestingly, this range of dark matter mass is generally compatible with the stringent limits of the Fermi-LAT gamma-ray observations of dwarf galaxies \citep{Ackermann,Albert,Cholis}. Nevertheless, all of the above claims are still controversial as the uncertainties of Galactic pulsars' contributions are significant \citep{Macias}. More observational data and analyses are required to settle the debate.

In this article, we target on a relatively high-redshift galaxy cluster, the A697 cluster (redshift $z=0.282$, distance $D=911$ Mpc), which has an almost spherical large diffuse radio halo at the central region. The radio continuum spectrum has been obtained by several radio observations. Besides, it is a very massive and bright galaxy cluster. The hot gas temperature profile is almost a constant without a cool core. These properties suggest that the A697 cluster is an ideal target object for constraining dark matter. Furthermore, it would be the first study of using a relatively high-redshift galaxy cluster for constraining dark matter, which can provide complementary information of dark matter properties.
 
\section{Dark matter annihilation model}
Dark matter annihilation would produce a large amount of high-energy electrons, positrons, photons and neutrinos. In particular, the high-energy electrons and positrons would interact with magnetic field to produce synchrotron radiations, which can be detected by radio telescopes. Many previous studies have examined the constraints of dark matter annihilation using radio data \citep{Colafrancesco,Egorov,Colafrancesco2,Chan3}. The spectra of these high-energy electrons and positrons for different annihilation channels are well-determined by numerical calculations \citep{Cirelli}. The synchrotron power at frequency $\nu$ is given by \citep{Storm}
\begin{equation}
P_{\rm syn}=\int_0^\pi d\theta \frac{(\sin \theta)^2}{2} 2\pi \sqrt{3}r_em_ec\nu_gF_{\rm syn} \left(\frac{x}{\sin \theta} \right),
\end{equation}
where $\nu_g=eB(r)/(2\pi m_ec)$, $B(r)$ is the magnetic field strength profile, $r_e$ is the classical electron radius, and the quantities $x$ and $F_{\rm syn}$ are defined as
\begin{equation}
x= \frac{2 \nu(1+z)}{3 \nu_g \gamma^2} \left[1+ \left(\frac{\gamma \nu_p}{\nu(1+z)} \right)^2 \right]^{3/2},
\end{equation}
where $\gamma$ is the Lorentz factor of the high-energy electrons or positrons and $\nu_p=8890[n(r)/1~{\rm cm}^{-3}]^{1/2}$ Hz is the plasma frequency, and
\begin{equation}
F_{\rm syn}(y)=y \int_y^{\infty} K_{5/3}(s)ds \approx 1.25y^{1/3}e^{-y}(648+y^2)^{1/12}.
\end{equation}

The high-energy electrons and positrons would cool down mainly via four processes: synchrotron radiation, inverse Compton scattering of the Cosmic Microwave Background photons, Bremsstrahlung radiation and Coulomb losses. The total cooling rate (in $10^{-16}$ GeV s$^{-1}$) of a high-energy electron or positron with energy $E$ is given by \citep{Colafrancesco}
\begin{equation}
\begin{aligned}
b(E)
=&0.0254E^2[B(r)]^2+0.25E^2(1+z)^4+1.51n(r)\left[0.36+\log \left(\frac{\gamma}{n(r)} \right) \right]
\\
& +6.13n(r) \left[1+\frac{1}{75} \log \left(\frac{\gamma}{n(r)} \right) \right],
\end{aligned}
\end{equation}
where $n(r)$, $E$ and $B(r)$ are in the units of cm$^{-3}$, GeV and $\mu$G respectively. The thermal electron number density profile in a galaxy cluster is usually modeled by the $\beta$-model \citep{Chen}
\begin{equation}
n(r)=n_0 \left(1+ \frac{r^2}{r_c^2} \right)^{-3\beta/2},
\end{equation}
where $n_0$ is the central number density, $r_c$ is the scale radius and $\beta$ is the index parameter. The $\beta$-model profile is obtained from the surface brightness profile observed by X-ray. It is usually constructed by azimuthal averaging in concentric bins in X-ray observations \citep{Reiprich,Chen}. Therefore, the resulting thermal electron number density is assumed spherically symmetric. In fact, there is another analytic functional form suggested by \citet{Bulbul} to describe the thermal electron number density profile. It is determined from the polytropic equation of state and this model has a total of 8 parameters \citep{Landry}. However, for our target galaxy cluster, both models can give almost the same goodness of fits for the observed surface brightness profile \citep{Landry}. Therefore, for simplicity, we follow the $\beta$-model to perform our analysis. Furthermore, we can see that the above formulation depends on the redshift $z$ of the galaxy cluster. In particular, the inverse Compton scattering (the second term in Eq.~(4)) of the high-energy electrons and positrons would be more significant for high-redshift galaxy clusters.

Besides, theoretical models suggest that the magnetic field strength profile in a galaxy cluster follows the thermal electron density profile \citep{Dolag,Govoni}: 
\begin{equation}
B(r)=B_0 \left[ \left(1+ \frac{r^2}{r_c^2} \right)^{-3\beta/2} \right]^{\eta},
\end{equation}
where $B_0$ is the central magnetic field strength and $\eta=0.5-1.0$ is the index modeled in simulations. The central magnetic field strength $B_0$ can be written in terms of the central thermal electron density $n_0$ and the central temperature of the hot gas $T_0$: $B_0 \propto \epsilon^{-1/2}n_0^{1/2}T_0^{3/4}$, with $\epsilon=0.5-1$ \citep{Govoni,Kunz}. The value of $B_0$ for a typical galaxy cluster is about 10 $\mu$G.

The cooling time scale of the high-energy electrons and positrons is much smaller than their diffusion scale. In other words, most of the high-energy electrons and positrons would cool down to non-relativistic (mainly via synchrotron cooling and inverse Compton scattering) before leaving the galaxy cluster. The diffusion length of a high-energy electron with initial energy $E$ can be approximately given by \citep{Yuan}
\begin{equation}
\lambda \sim 79~{\rm kpc} \left(\frac{D_0}{10^{30}~{\rm cm^3}} \right)^{1/2} \left(\frac{\omega_0}{\rm 1~eV/cm^3}\right)^{-1/2} \left(\frac{E}{1~\rm GeV} \right)^{-1/3},
\end{equation}
where $\omega_0$ is the total radiation energy density and $D_0$ is the diffusion coefficient. Considering the outskirt regions of a galaxy cluster, the synchrotron radiation with $B \sim 1$ $\mu$G and inverse Compton scattering give $\omega_0 \sim 0.65$ eV/cm$^3$. By taking a conservative diffusion coefficient $D_0 \sim 10^{30}$ cm$^2$ s$^{-1}$, the diffusion length in outskirt regions is less than 100 kpc for $E \sim 1$ GeV, which is much smaller than the typical size of a galaxy cluster ($\sim 1$ Mpc). The diffusion length would be much smaller at the central region as the magnetic field strength is much larger ($\omega_0 \propto B^2$ for synchrotron radiation). Therefore, in general, the diffusion process is insignificant compared with the cooling and we can neglect the diffusion term in the diffusion equation. The equilibrium high-energy electron or positron number density energy spectrum is thus given by \citep{Storm}
\begin{equation}
\frac{dn_e}{dE}=\frac{(\sigma v)[\rho_{DM}(r)]^2}{2m^2b(E)} \int_E^m \frac{dN_{e,inj}}{dE'}dE',
\end{equation}
where $\rho_{DM}(r)$ is the dark matter density profile and $dN_{e,inj}/dE'$ is the injection energy spectrum of dark matter annihilation. 

Assuming the dark matter content dominates the whole galaxy cluster and the dark matter distribution is spherically symmetric. The dark matter density can be obtained by assuming the hot gas in hydrostatic equilibrium: 
\begin{equation}
\rho_{DM}(r)=\frac{1}{4 \pi r^2} \frac{d}{dr} \left[-\frac{kTr}{\mu m_pG} \left(\frac{d \ln n(r)}{d \ln r}+ \frac{d \ln T}{d \ln r} \right) \right],
\end{equation}
where $\mu=0.59$ is the molecular weight and $m_p$ is the proton mass. If $T$ is almost a constant in the galaxy cluster, using Eq.~(5), Eq.~(9) can be simplified to 
\begin{equation}
\rho_{DM}(r)=\frac{3 \beta kT}{4 \pi G \mu m_p} \left[ \frac{r^2+3r_c^2}{(r^2+r_c^2)^2} \right].
\end{equation}
Beside the hydrostatic density profile, we also use the Navarro-Frenk-White (NFW) profile \citep{Navarro} to model the dark matter density profile.

Combining the above equations, the radio flux density emitted from a galaxy cluster due to dark matter annihilation is:
\begin{equation}
S_{\rm DM}(\nu)=\frac{1}{4\pi D^2} \int_0^R \int_{m_e}^m 2 \frac{dn_e}{dE}P_{\rm syn} dE(4\pi r^2)dr,
\end{equation}
where $D$ is the distance to the galaxy cluster and $R \approx 2.5$ Mpc is the virial radius. The factor 2 in the above equation indicates the contributions of both high-energy electrons and positrons. Furthermore, simulations show that sub-structures in galaxy clusters can enhance the annihilation rate by a factor $(1+B_{\rm sub})$ \citep{Gao,Anderhalden,Marchegiani,Sanchez}. The redshift-dependent boost factor $B_{\rm sub}$ can be represented by an empirical functional form in terms of the total halo mass $M_{\rm tot}$ and the redshift $z$ \citep{Ando}:
\begin{equation}
\log B_{\rm sub}= \frac{X}{1+\exp[-a'(\log M_{\rm tot}-m_1)]}+c' \left[1+ \frac{Y}{1+\exp[-b'(\log M_{\rm tot}-m_2)]} \right],
\end{equation}
where $X=2.2e^{-0.75z}+0.67$, $Y=2.5e^{-0.005z}+0.8$, $a'=0.1e^{-0.5z}+0.22$, $b'=0.8e^{-0.5(z-12)^4}-0.24$, $c'=-0.0005z^3-0.032z^2+0.28z-1.12$, $m_1=-2.6z+8.2$ and $m_2=0.1e^{-3z}-12$. Here, the empirical functional form and the parameters used follow the mass-concentration model in \citet{Okoli}. The resulting boost factor for galaxy clusters at present time ($B_{\rm sub} \sim 30$ at $z=0$) is consistent with other simulation results \citep{Sanchez}. In fact, there is another parametric form of boost factor shown in \citet{Ando} which assumes the concentration function obtained in \citet{Correa}. However, the resulting boost factor at $z=0$ is $B_{\rm sub} \approx 3$, which is too small compared with the standard range $B_{\rm sub} \sim 30-35$ for galaxy clusters \citep{Sanchez,Marchegiani}. In the followings, we will first use the functional form in Eq.~(12) (i.e. assuming the mass-concentration model in \citet{Okoli}) to do the analysis. After that, we will also consider the functional form assuming the concentration function in \citet{Correa} as the possible lower limit of boost factor and a reasonably larger boost factor as the upper limit (see below).
 
\section{Data fitting}
We use the archival radio continuum spectral data of the A697 cluster for analysis \citep{Macario}. For the A697 cluster, the temperature profile is almost constant up to at least 800 kpc (see Fig.~1). Since majority of the dark matter annihilation contribution on radio flux is found within 800 kpc, the assumption of constant $T$ used in Eq.~(10) is a very good approximation. 

Apart from the possible contribution of dark matter annihilation, the diffuse background cosmic rays in the A697 cluster also contribute a significant amount of the observed total radio signals. The A697 cluster has a giant radio halo near the center which emits strong diffuse radio signals \citep{Macario}. Moreover, it does not have a clear radio relic in cluster peripheral regions \citep{Feretti,vanWeeren}. Therefore, the diffuse radio emissions due to cosmic rays are quite centralized and this property suggests that the A697 cluster is an excellent target to get the radio constraints of dark matter. 

To account for the diffuse background radio emissions, we write the predicted total radio signals as $S_{\rm tot}=S_{\rm CR}+S_{\rm DM}$, where $S_{\rm CR}$ is the radio contribution from the cosmic rays. The spectral shapes of the cosmic rays are model-dependent. Several models have been proposed to account for the radio spectral shapes of diffuse cosmic rays, including primary electron emission models \citep{Jaffe,Rephaeli,Rephaeli2}, secondary electron emission models \citep{Dennison} and the in-situ acceleration models \citep{Jaffe,Roland}. These models could be parametrized by the following three forms \citep{Thierbach}: 
\begin{equation}
S_{\rm CR}=S_{\rm CR,0} \left(\frac{\nu}{\rm GHz} \right)^{-\alpha} \left[\frac{1}{1+(\nu/\nu_s)^{\Gamma}} \right],
\end{equation}
where $\Gamma=0.5$ or 1,
\begin{equation}
S_{\rm CR}=S_{\rm CR,0} \left( \frac{\nu}{\rm GHz} \right)^{-\alpha},
\end{equation}
or
\begin{equation}
S_{\rm CR}=S_{\rm CR,0} \left(\frac{\nu}{\rm GHz} \right)^{-\alpha} \exp(-\nu^{1/2}/\nu_s^{1/2}).
\end{equation}
We denote the above three model forms by M1 (Eq.~(13)), M2 (Eq.~(14)) and M3 (Eq.~(15)) respectively. In the above three parametric forms, $S_{\rm CR,0}$, $\alpha$ and $\nu_s$ are free parameters (without any constraints) for fitting the observed radio spectrum \citep{Thierbach}. The above three parametric forms have been used to examine the radio spectral features of the giant radio halo of the Coma cluster (the Coma C) and some good fits have been obtained \citep{Thierbach}. Among these parametric forms, the model M3 gives the best fit for the Coma C radio spectrum \citep{Thierbach}. In fact, the Coma C and the giant radio halo in the A697 cluster belong to the same classification of the radio halos \citep{vanWeeren}. Therefore, we expect that the above three models can also give some good fits for the cosmic-ray component. Also, these three models together with the three completely free parameters can almost represent the major spectral shapes and emission models for giant radio halos, although they might not be exhaustive. 

For the A697 cluster, the values of the hot gas parameters are $\beta=0.58^{+0.04}_{-0.07}$, $r_c=192^{+27}_{-29}$ kpc and $n_0=0.0090 \pm 0.0007$ cm$^{-3}$ (assumed the Hubble parameter $h=0.70$) \citep{Landry}. For the central temperature $T_0=10.2$ keV obtained in the Chandra observations \citep{Cavagnolo}, the possible range of the central magnetic field strength is $B_0=11.0-15.7$ $\mu$G. For the NFW profile, we obtain the corresponding parameters from the dynamical data of the A697 cluster \citep{Girardi}. Using all of the above parameters, we can obtain the magnetic field profile $B(r)$, dark matter density profiles $\rho_{DM}(r)$ (hydrostatic and NFW) and the total dark matter halo mass $M_{\rm tot}=1.66 \times 10^{15}M_{\odot}$. The boost factor of the A697 cluster at $z=0.282$ is $B_{\rm sub}=15.6$ for the mass-concentration model in \citet{Okoli} (i.e. Eq.~(12)). We take $B_{\rm sub}=3.82$ at $z=0.282$ using the concentration function in \citet{Correa} as our lower limit of the boost factor and we arbitrarily set $B_{\rm sub}=31.2$ (two times of our benchmark value) to be the upper limit.

By using Eq.~(11), we can predict the radio flux contributed by dark matter annihilation $S_{\rm DM}$ as a function of radio frequencies $\nu$. To minimize the number of free parameters, we take the thermal annihilation cross section $\sigma v=2.2 \times 10^{-26}$ cm$^3$ s$^{-1}$ predicted by standard cosmology \citep{Steigman}. Therefore, only one free parameter $m$ in $S_{\rm DM}$ and two to three free parameters ($S_{\rm CR,0}$, $\alpha$ and $\nu_s$) in $S_{\rm CR}$ are involved in the analysis. 

We compare the predicted $S_{\rm tot}$ with the observed radio flux spectrum $S_{\rm obs}$ of the A697 cluster for different annihilation channels and cosmic-ray parametric forms. The goodness of fits can be calculated by the reduced $\chi^2$ value:
\begin{equation} 
\chi_{\rm red}^2=\frac{1}{N} \sum_i \frac{(S_{\rm tot,i}-S_{\rm obs,i})^2}{\sigma_{\rm obs,i}^2},  
\end{equation}
where $N$ is the number of the degrees of freedom, $\sigma_{\rm obs,i}$ is the uncertainties of the observed radio flux density. For each $m$ and annihilation channel, we tune the parameters such that the value of $\chi_{\rm red}^2$ is minimized. Then we plot the minimum $\chi_{\rm red}^2$ as a function of $m$ for each annihilation channel and cosmic-ray parametric form (see Fig.~2 and Fig.~3).  

Among the three parametric forms of cosmic-ray emission, we find that only the M3 model form in Eq.~(15) plus dark matter annihilating via the leptophilic channels ($e^+e^-$, $\mu^+\mu^-$ and $\tau^+\tau^-$) can give the values $\chi_{\rm red}^2<0.7$, which mean very good fits. In general, using the hydrostatic density profile gives much better fits (smaller $\chi_{\rm red}^2$) than using the NFW profile for all of the three cosmic-ray models and also for the five annihilation channels we considered (see Fig.~2). Therefore, in the followings, we mainly focus on the results assuming the hydrostatic density profile. In Fig.~3, we show the effects of the magnetic field strength on the goodness of fits. Generally speaking, assuming $B_0=15.7$ $\mu$G and $\eta=0.5$ give better fits for the five annihilation channels. 

Using our benchmark value $B_{\rm sub}=15.6$ and neglecting the uncertainties of parameters, the best-fit $m$ for $e^+e^-$, $\mu^+\mu^-$ and $\tau^+\tau^-$ channels are 60 GeV, 40 GeV and 20 GeV respectively for the M3 model (see Table 1). The corresponding $2\sigma$ ranges of $m$ are 48-130 GeV, 31-80 GeV and 10-54 GeV. However, the overall fits for the $b\bar{b}$ and $W^+W^-$ channels are not very good ($\chi_{\rm red}^2>3.3$) for this best-fit scenario (see Fig.~3), which are ruled out by $2.1 \sigma$. For $m$ becomes very large ($m>200$ GeV), the contribution of dark matter would be less significant and the minimum $\chi_{\rm red}^2$ would approach a constant value 4.4 (see Fig.~3). This represents the fits for $S_{\rm tot}=S_{\rm CR}$ without dark matter annihilation (the null hypothesis). Therefore, by comparing the smallest value $\chi_{\rm red}^2 \sim 0.5$ with the asymptotic value $\chi_{\rm red}^2 \approx 4.4$ for the null hypothesis, the statistical significance of having dark matter annihilation is about $2.7 \sigma$, which indicates a strong signal of the existence of dark matter annihilation. 

For the M1 or M2 model form plus dark matter annihilation, the smallest reduced $\chi^2$ values are $\chi_{\rm red}^2 \approx 1$, which means the overall fits are still good. However, the asymptotic values of $\chi_{\rm red}^2$ in the large $m$ regimes for M1 and M2 model forms are 2 and 1.4 respectively (see Fig.~3). It means that even for no dark matter contribution, the M1 and M2 cosmic-ray model forms can still give good fits. The statistical significance of having dark matter annihilation in these two models is less than $1.4 \sigma$, which indicates barely positive signals of dark matter annihilation. Overall speaking, although we cannot rule out the possibility of the M1 or M2 model without dark matter annihilation, the M3 model plus dark matter annihilation can give the best fits and it reveals a strong signal of dark matter annihilation. In Fig.~4, we show the best spectral fits for 4 popular annihilation channels using the M3 model and the corresponding components $S_{\rm DM}$ and $S_{\rm CR}$. 

Moreover, we also examine the impact of our results due to the uncertainties of the parameters used. We vary the parameters $n_0=0.0090 \pm 0.0007$ cm$^{-3}$, $\beta=0.58^{+0.04}_{-0.07}$, $r_c=192^{+27}_{-29}$ kpc \citep{Landry} and $T=10^{+2}_{-4}$ keV \citep{Cavagnolo} within their $1\sigma$ ranges such that they can give a possible range of the dark matter density. We also consider a large range of possible boost factor $B_{\rm sub}=3.82-31.2$. In Fig.~5 and Fig.~6, we show the effects of the goodness of fits due to the uncertainties of the parameters (assuming the hydrostatic density profile). We can see that the major effects of the uncertainties are the best-fit ranges of dark matter mass $m$. The possible ranges of $m$ is somewhat larger if uncertainties are taken into account. For the best-fit scenario (the M3 model) considering all uncertainties of parameters and boost factor, the $2\sigma$ ranges of $m$ become 15-250 GeV, 10-150 GeV, 10-150 GeV, 15-150 GeV and 90-100 GeV respectively for the $e^+e^-$, $\mu^+\mu^-$, $\tau^+\tau^-$, $b{\bar{b}}$ and $W^+W^-$ channels. For the M1 and M2 models, the goodness of fits (the values of $\chi_{\rm red}^2$) do not have big changes. Nevertheless, for the M3 model, the uncertainties of parameters have some effects on the goodness of fits. The upper and lower limits of the dark matter density give slightly larger $\chi_{\rm red}^2$ values (less good fits) for the leptophilic annihilation channels. Similarly, for the boost factor, the major effect of a larger or smaller boost factor is the best-fit values of $m$. Moreover, the two extreme values of $B_{\rm sub}=3.82$ and $B_{\rm sub}=31.2$ also give slightly larger $\chi_{\rm red}^2$ values, except for a few cases (see Fig.~6). In particular, in the large boost factor plus large dark matter density regime and assuming the M3 model, the statistical significance of having dark matter annihilation has increased to $(2-2.5)\sigma$ for the $b\bar{b}$ and $W^+W^-$ channels (see Fig.~7). Therefore, a relatively larger dark matter contribution can manifest a larger signal for these two channels. However, this effect cannot be seen in the M1 and M2 models. Generally speaking, we still get a strong signal of dark matter annihilation for the M3 model (more than $2 \sigma$ statistical significance) after considering the uncertainties of the parameters. However, the possible ranges of $m$ for different annihilation channels are somewhat larger.

\section{Discussion}
In this article, we target on a relatively high-redshift galaxy cluster A697 and constrain the dark matter properties by its radio continuum spectrum. We have obtained very good fits for three leptophilic channels ($e^+e^-$, $\mu^+\mu^-$ and $\tau^+\tau^-$), assuming the M3 model form. The statistical significance of the dark matter annihilation signal is about $2.7 \sigma$, which indicates a strong signal. However, we cannot completely rule out the possibility of the M1 and M2 models without dark matter annihilation. 

Including large ranges of uncertainties of parameters and boost factor, the best-fit $2\sigma$ ranges of dark matter mass $m$ for $e^+e^-$, $\mu^+\mu^-$ and $\tau^+\tau^-$ channels are 15-250 GeV, 10-150 GeV and 10-150 GeV respectively, assuming the thermal annihilation cross section and the hydrostatic density profile. Large ranges of the best-fit $m$ for $e^+e^-$ and $\mu^+\mu^-$ channels can satisfy the current gamma-ray limits \citep{Ackermann} and the AMS limits \citep{Cavasonza}. However, a considerable range of $m$ for $\tau^+\tau^-$ channel is in some tension with the new gamma-ray constraints \citep{Albert}. On the other hand, a recent study considering the inverse Compton scattering suggests that dark matter annihilating via the $\mu^+\mu^-$ channel with $m=57.4^{+4.6}_{-4.1}$ GeV can also account for the gamma-ray excess in our Galaxy \citep{Calore}. This range coincides with our results and it suggests that $\mu^+\mu^-$ channel may be another possible annihilation channel that we should pay more attention to.

Most of the previous studies show that $b\bar{b}$ channel can provide good fits to gamma-ray and radio data (best-fit $m \sim 30-80$ GeV) \citep{Daylan,Calore,Cholis,Brown,Brown2,Chan,Chan2}. However, in our best-fit scenario using the M3 model, the fits for the $b\bar{b}$ channel is not very good unless we consider the large boost factor and dark matter density regime. For this particular regime, the best-fit $2\sigma$ range of $m$ is 30-150 GeV, which is still consistent with the best-fit ranges based on gamma-ray and radio data. Note that our results are just based on the radio data of a single galaxy cluster. Furthermore, we need to pay attention to some possible systematic uncertainties involved in our analysis. For example, the systematic uncertainties of the magnetic field strength profile used may be significant. Moreover, we can see that we could not have a crystal-clear conclusion for the M1 and M2 cosmic-ray models. The $\chi_{\rm red}^2$ for these two models are quite small so that they are still regarded as good fits. Therefore, the constraints for the $b\bar{b}$ channel are still very weak due to the uncertainties. In fact, the frequency range of the data used ($\nu \approx 0.1-1.7$ GHz) is not wide enough to differentiate the impacts of different cosmic-ray models. In Fig.~4, we can see that the effect of dark matter annihilation on radio spectrum dominates at high frequency ($\nu \ge 1$ GHz) while the effect of cosmic rays dominates at low frequency ($\nu \sim 0.1$ GHz). If we can have the data with a larger range of frequency (e.g. $\nu=0.1-10$ GHz), the effects of different cosmic-ray models would be more differentiable and we can observe the specific impact of dark matter annihilation on the radio spectrum more easily. Therefore, further radio observations of the A697 cluster using higher frequencies (e.g. $\nu \sim 10$ GHz) are required to verify our results. 

Beside the uncertainties of the magnetic field profile and the cosmic-ray models, the uncertainties of the parameters used and the dark matter density profile also affect our results. First of all, we have found that using the hydrostatic density profile can give better fits than using the NFW density profile. In fact, the hydrostatic density profile may be more representative because it is directly derived from the baryonic matter (hot gas) distribution of the A697 cluster. The uncertainty of the hydrostatic equilibrium assumption is only 15-20\% \citep{Biffi}. However, for the NFW profile, the parameters used are constrained by the enclosed mass data assuming dynamical equilibrium \citep{Girardi}, which have more than 30\% error in mass estimation. For the consideration of the parameter uncertainties, the major effect is that we obtain a larger possible ranges of $m$. Nevertheless, very good fits with $\chi_{\rm red}^2<1$ for the M3 model are still obtained by varying the hot gas parameters within their $1\sigma$ uncertainties or the boost factor by a certain factor.

\begin{figure}
\vskip 10mm
 \includegraphics[width=140mm]{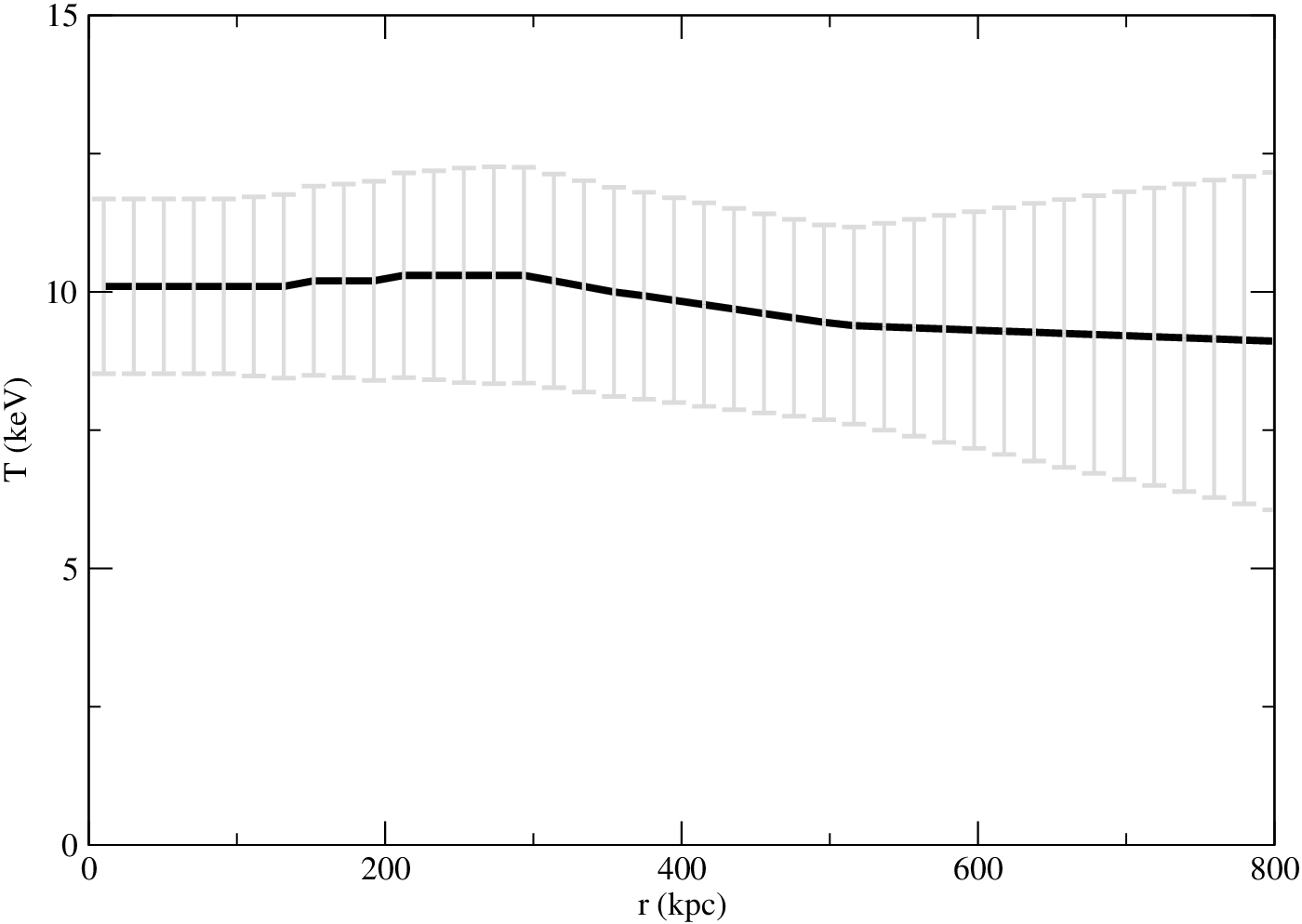}
 \caption{The temperature profile of the hot gas in A697 cluster (in keV). The data of the hot gas are extracted from the Chandra observations \citep{Cavagnolo}.}
\vskip 10mm
\end{figure}

\begin{figure}
\vskip 10mm
 \includegraphics[width=70mm]{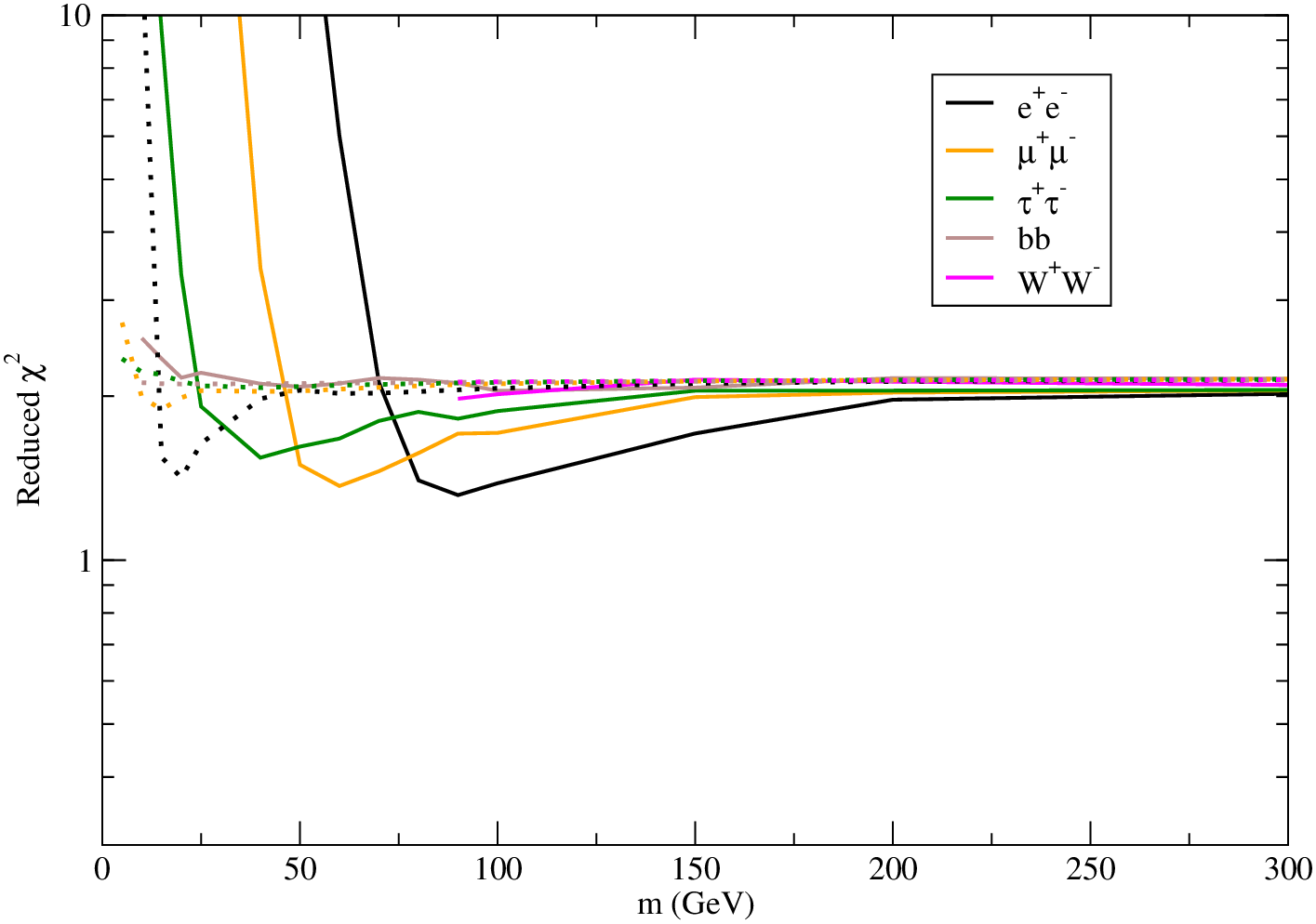}
~~ \includegraphics[width=70mm]{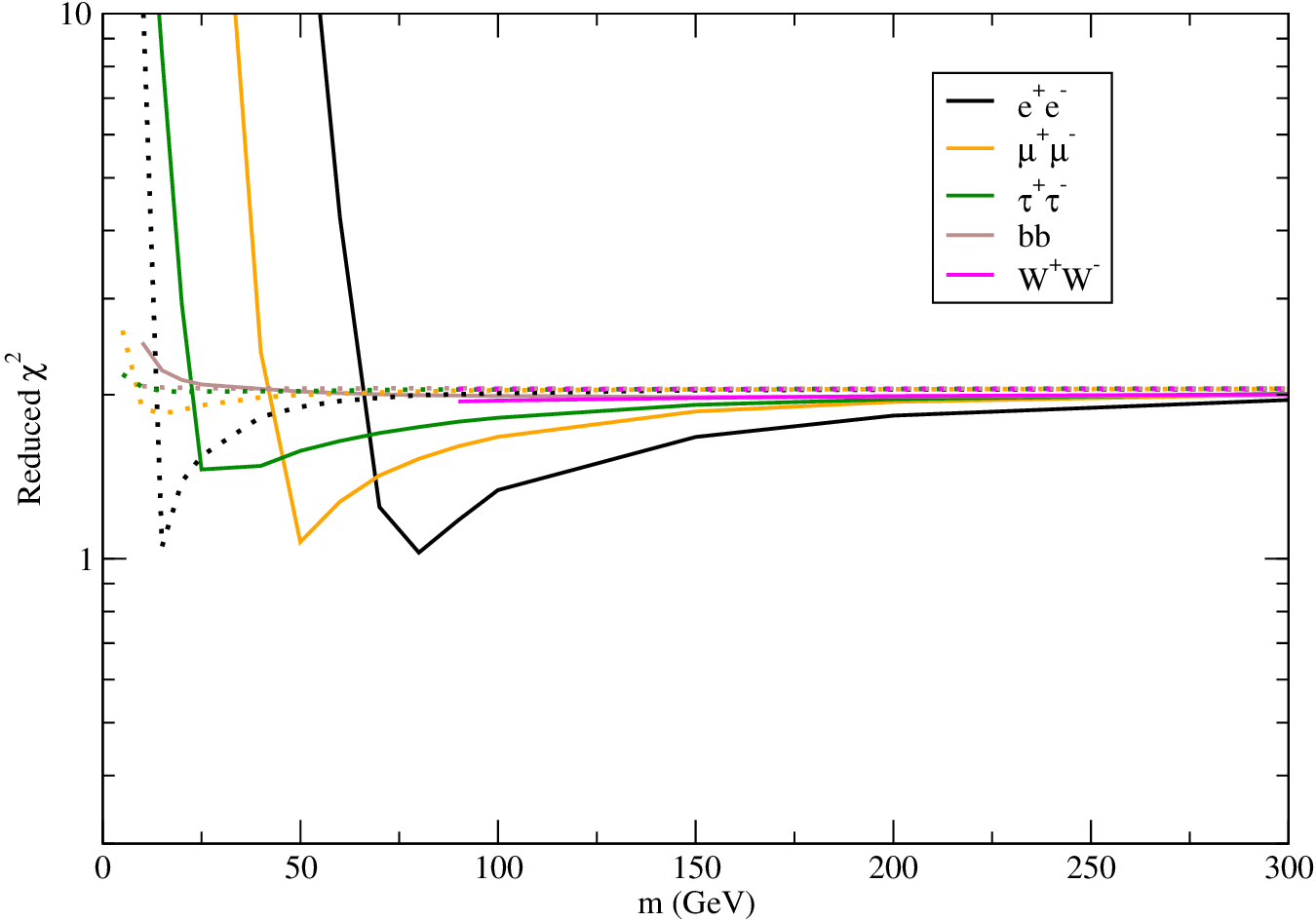} \newline
\newline
\newline
 \includegraphics[width=70mm]{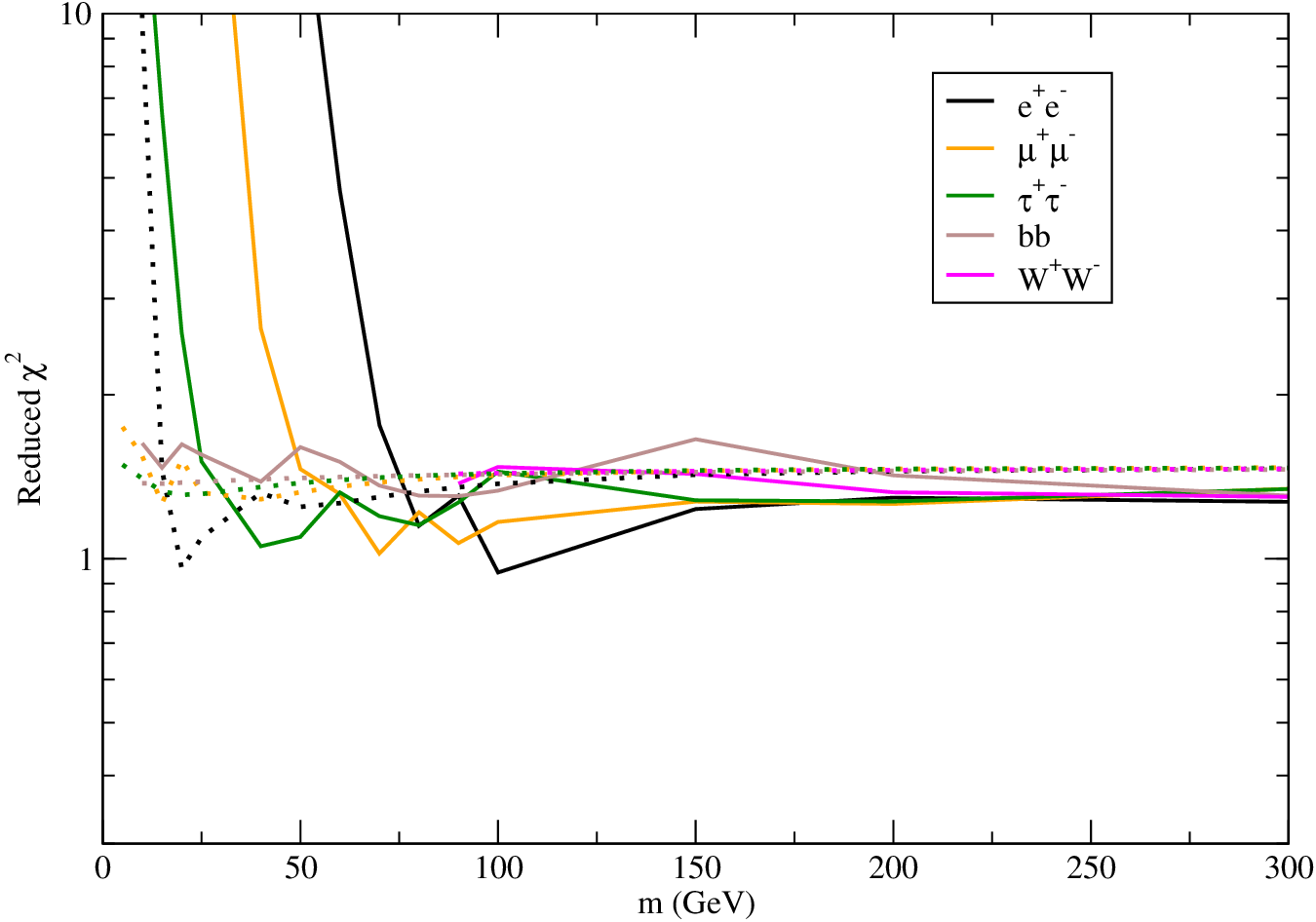}
~~ \includegraphics[width=70mm]{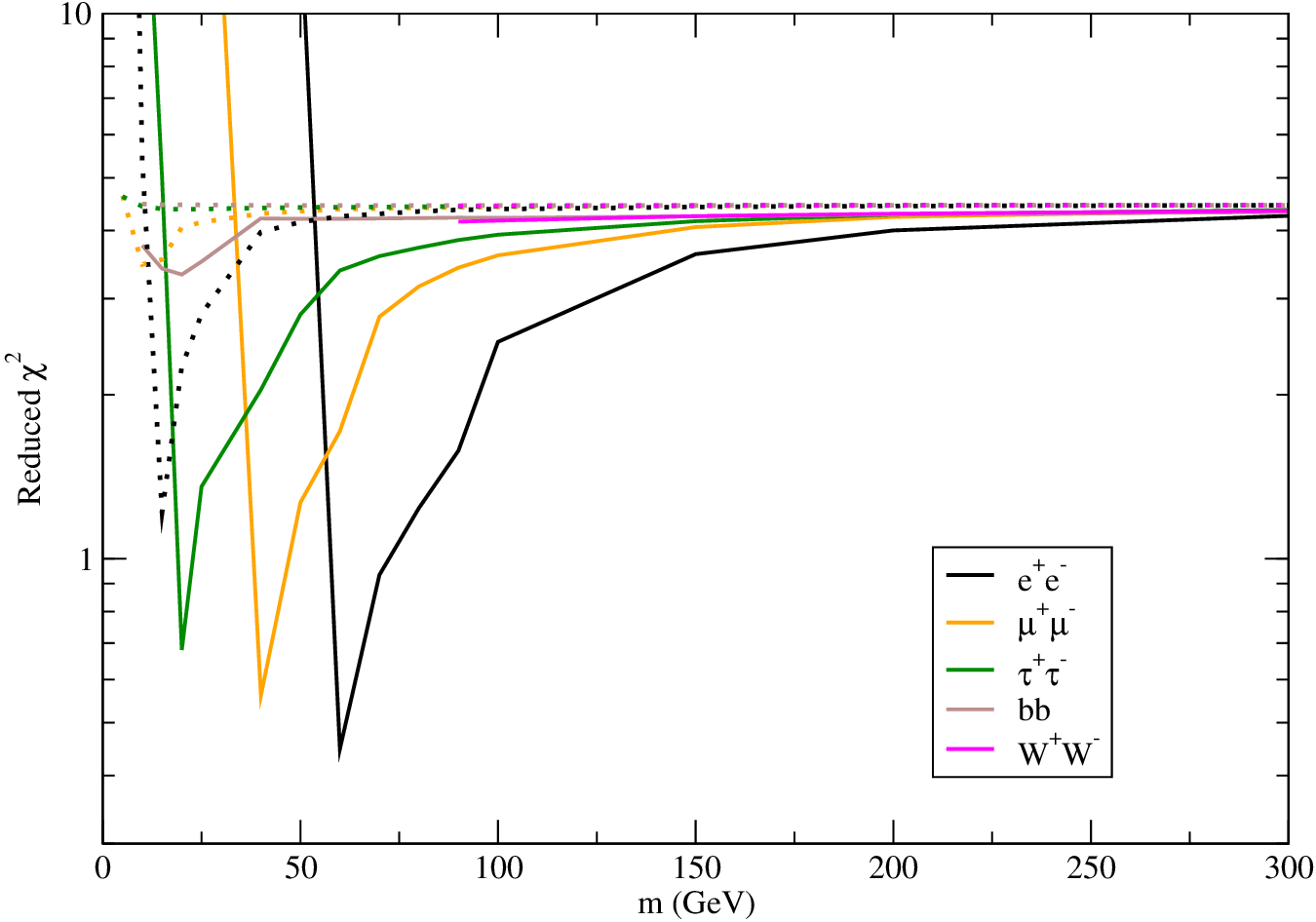}
 \caption{The minimum reduced $\chi^2$ values ($\chi_{\rm red}^2$) as a function of $m$ for five different annihilation channels (top left: M1 model with $\Gamma=0.5$; top right: M1 model with $\Gamma=1$; bottom left: M2 model; bottom right: M3 model). The dotted lines represent the fits with the NFW density profile and the solid lines represent the fits with the hydrostatic density profile. Here, we have assumed $B_0=15.7$ $\mu$G and $\eta=0.5$.}
\vskip 10mm
\end{figure}

\begin{figure}
\vskip 10mm
 \includegraphics[width=70mm]{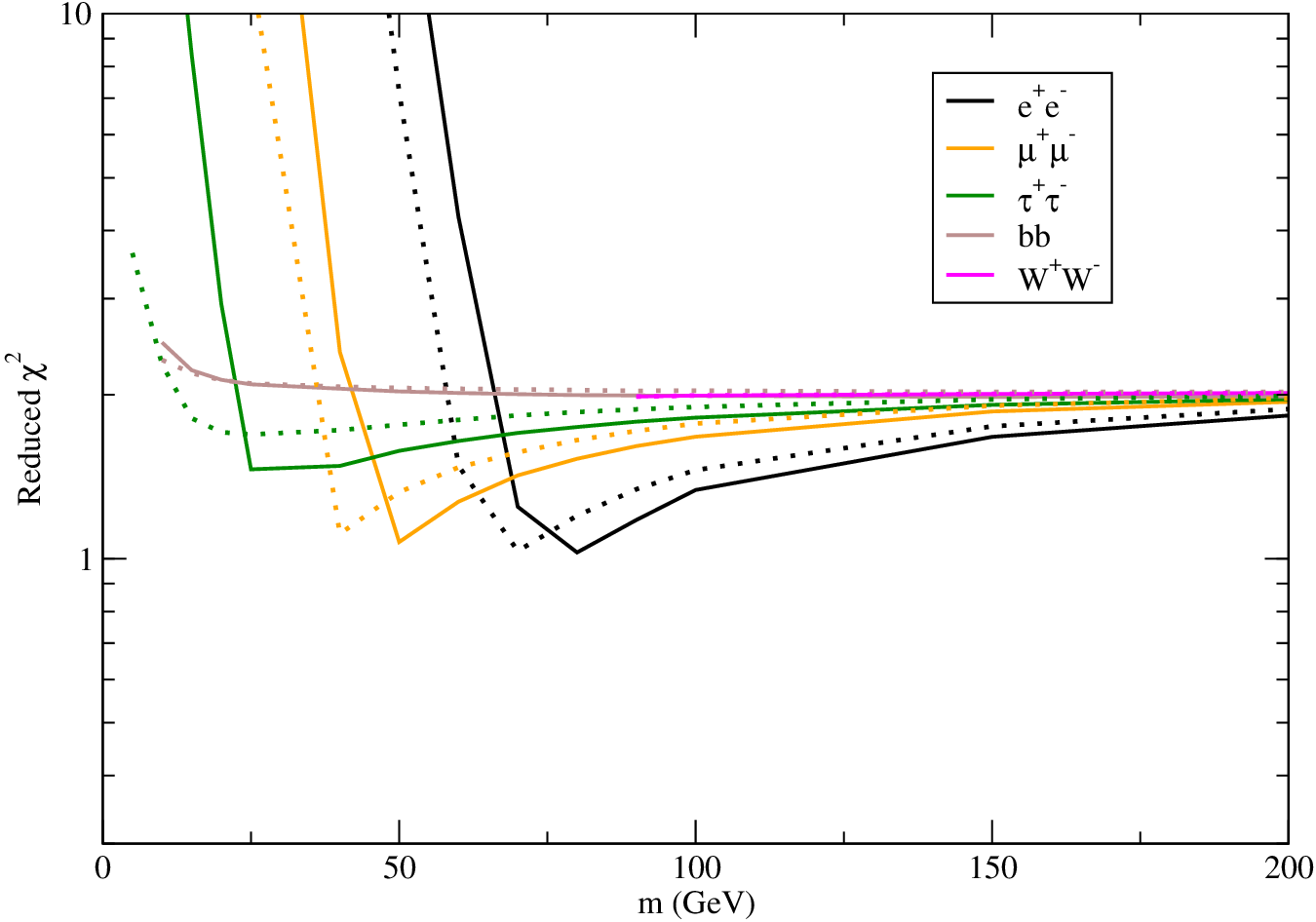}
~~ \includegraphics[width=70mm]{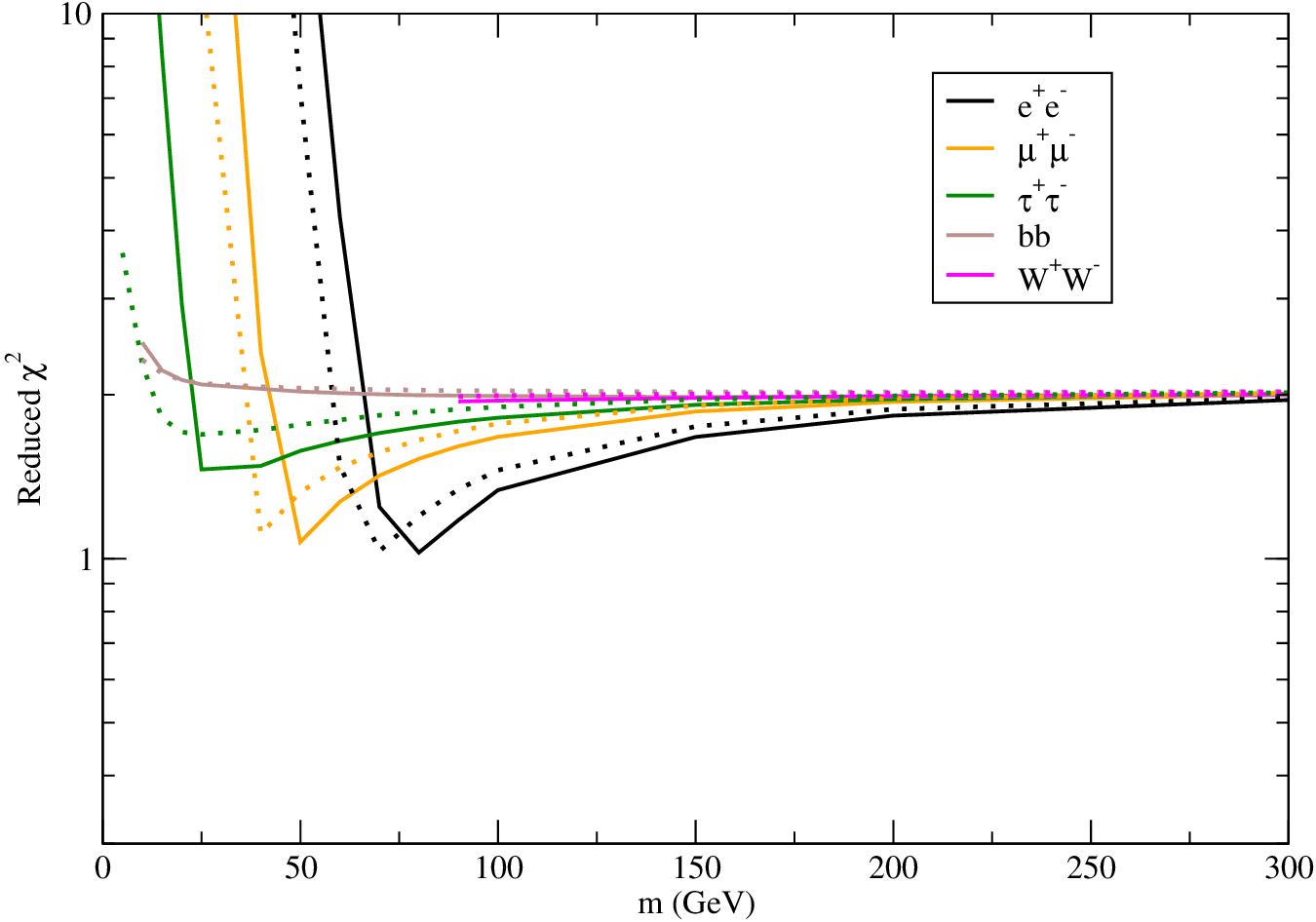} \newline
\newline
\newline
 \includegraphics[width=70mm]{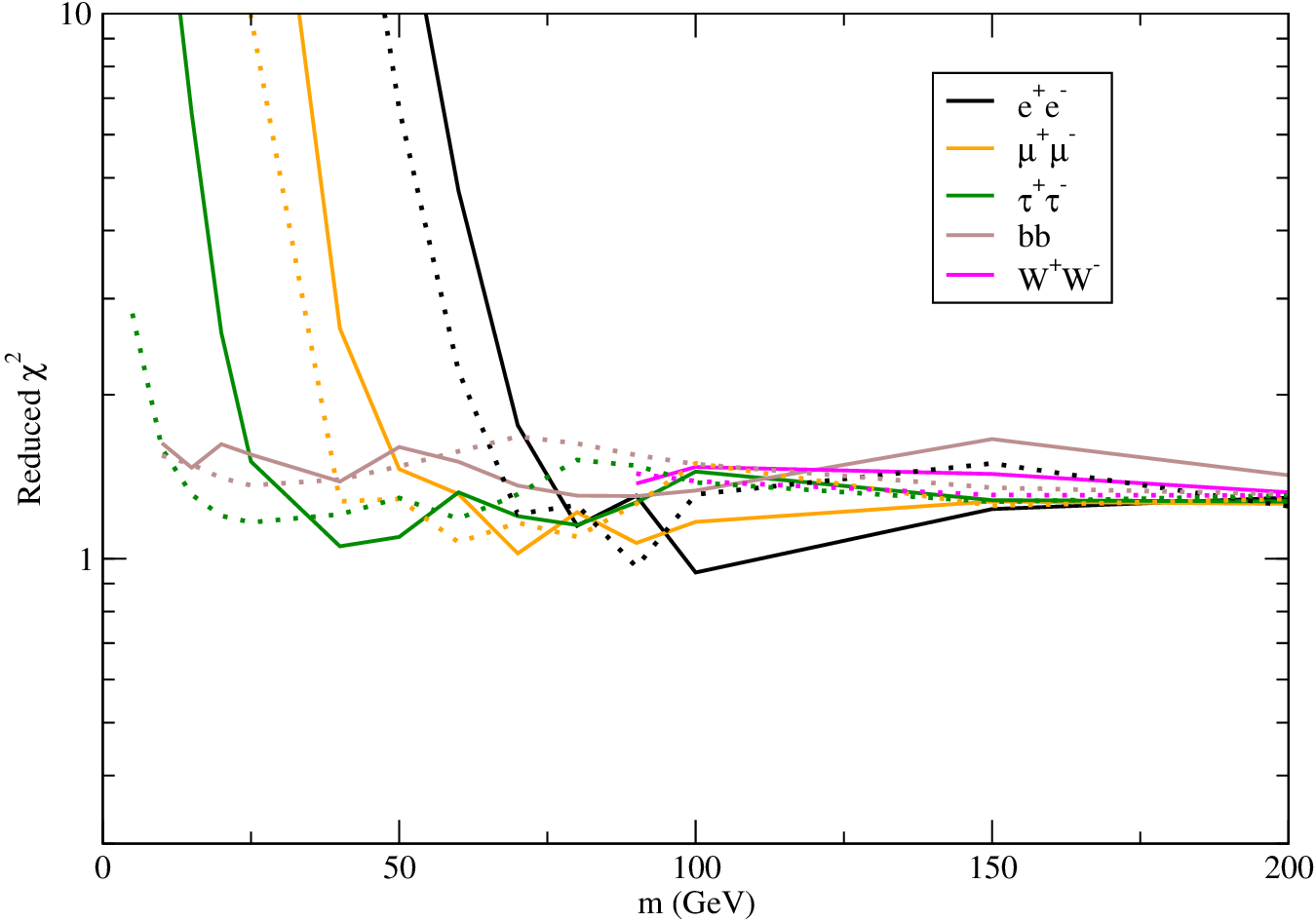}
~~ \includegraphics[width=70mm]{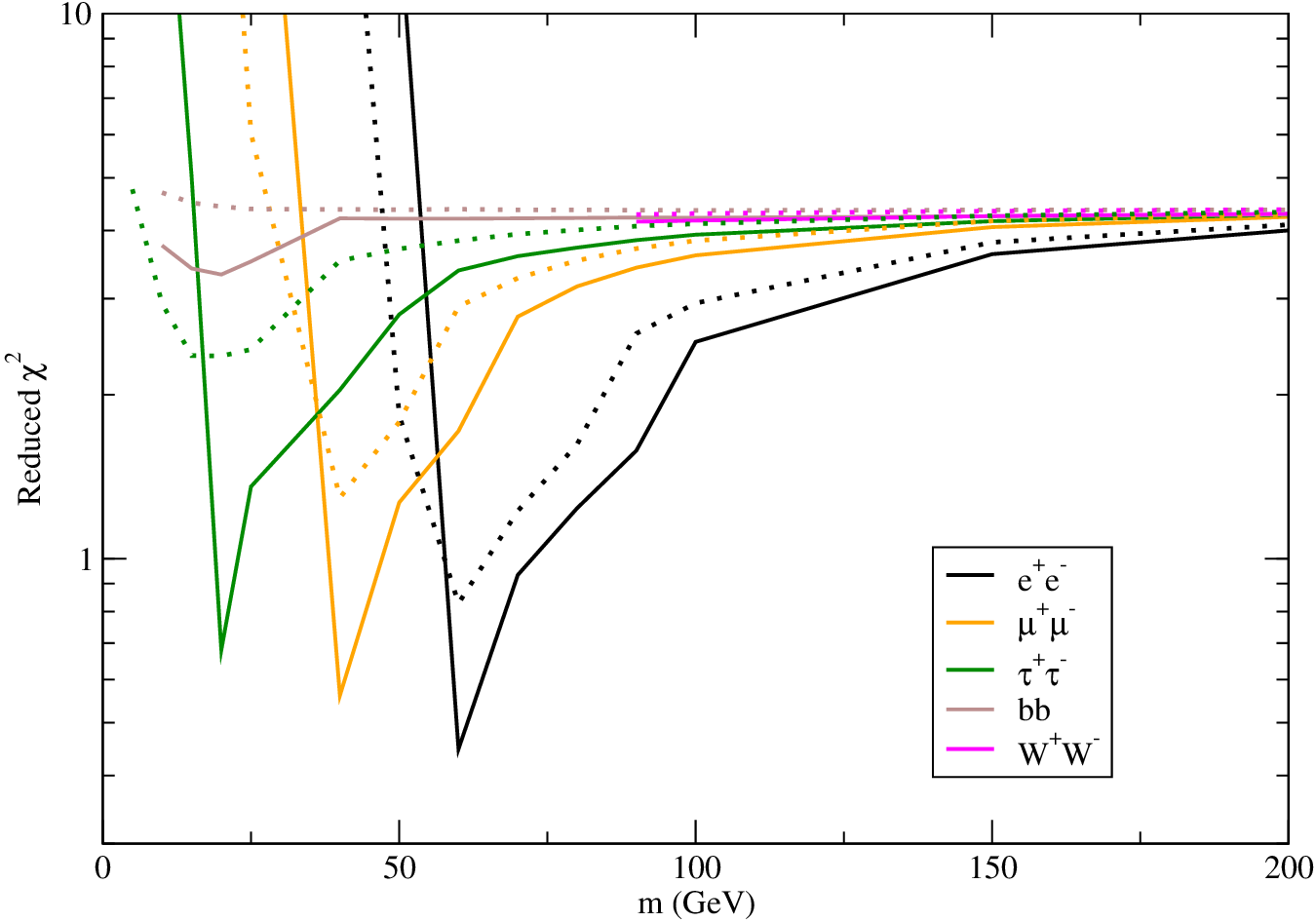}
 \caption{The minimum reduced $\chi^2$ values ($\chi_{\rm red}^2$) as a function of $m$ for five different annihilation channels (top left: M1 model with $\Gamma=0.5$; top right: M1 model with $\Gamma=1$; bottom left: M2 model; bottom right: M3 model). The dotted lines represent the fits with $B_0=11$ $\mu$G and $\eta=1$ and the solid lines represent the fits with $B_0=15.7$ $\mu$G and $\eta=0.5$.}
\vskip 10mm
\end{figure}

\begin{figure}
\vskip 10mm
 \includegraphics[width=70mm]{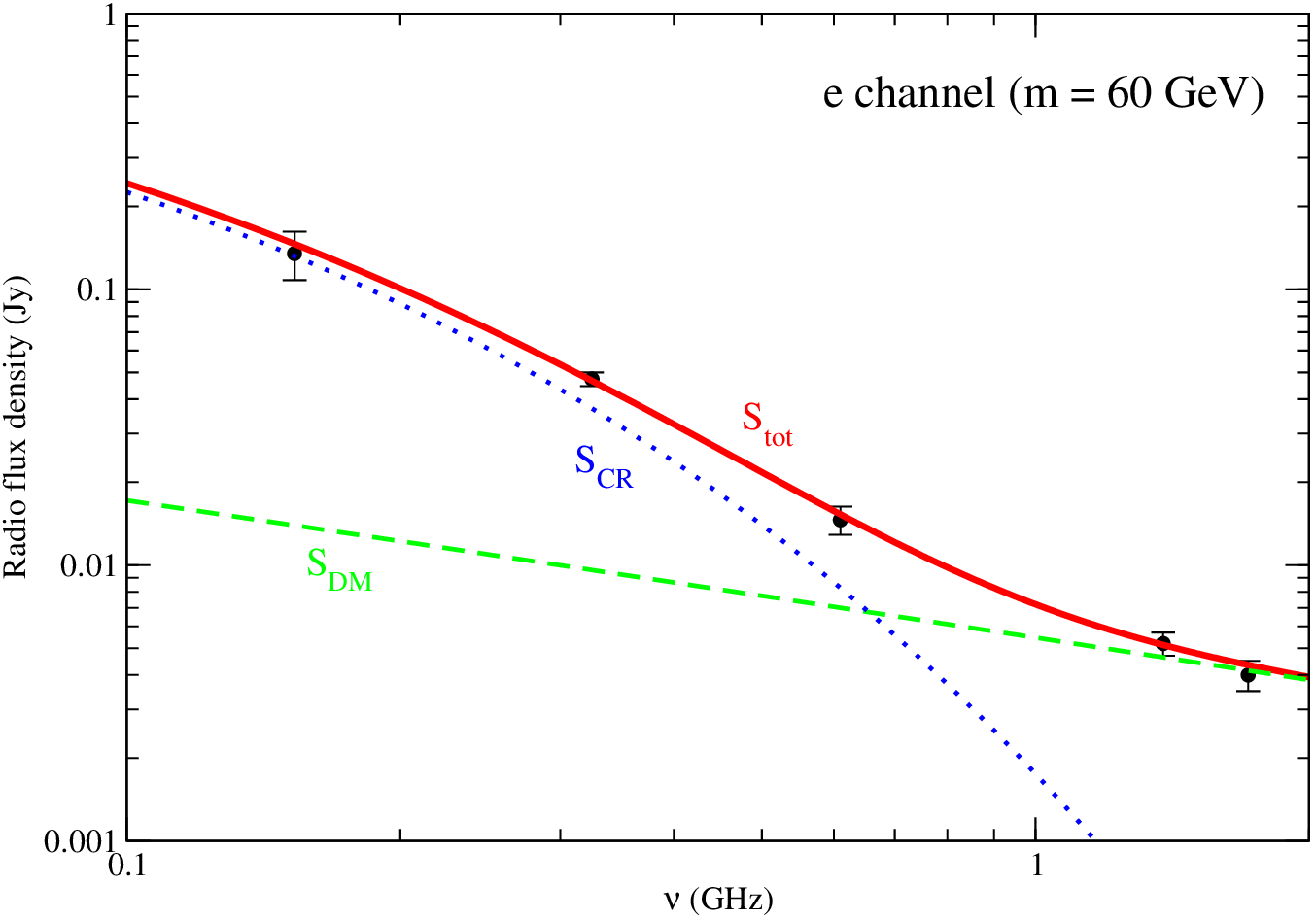}
~~ \includegraphics[width=70mm]{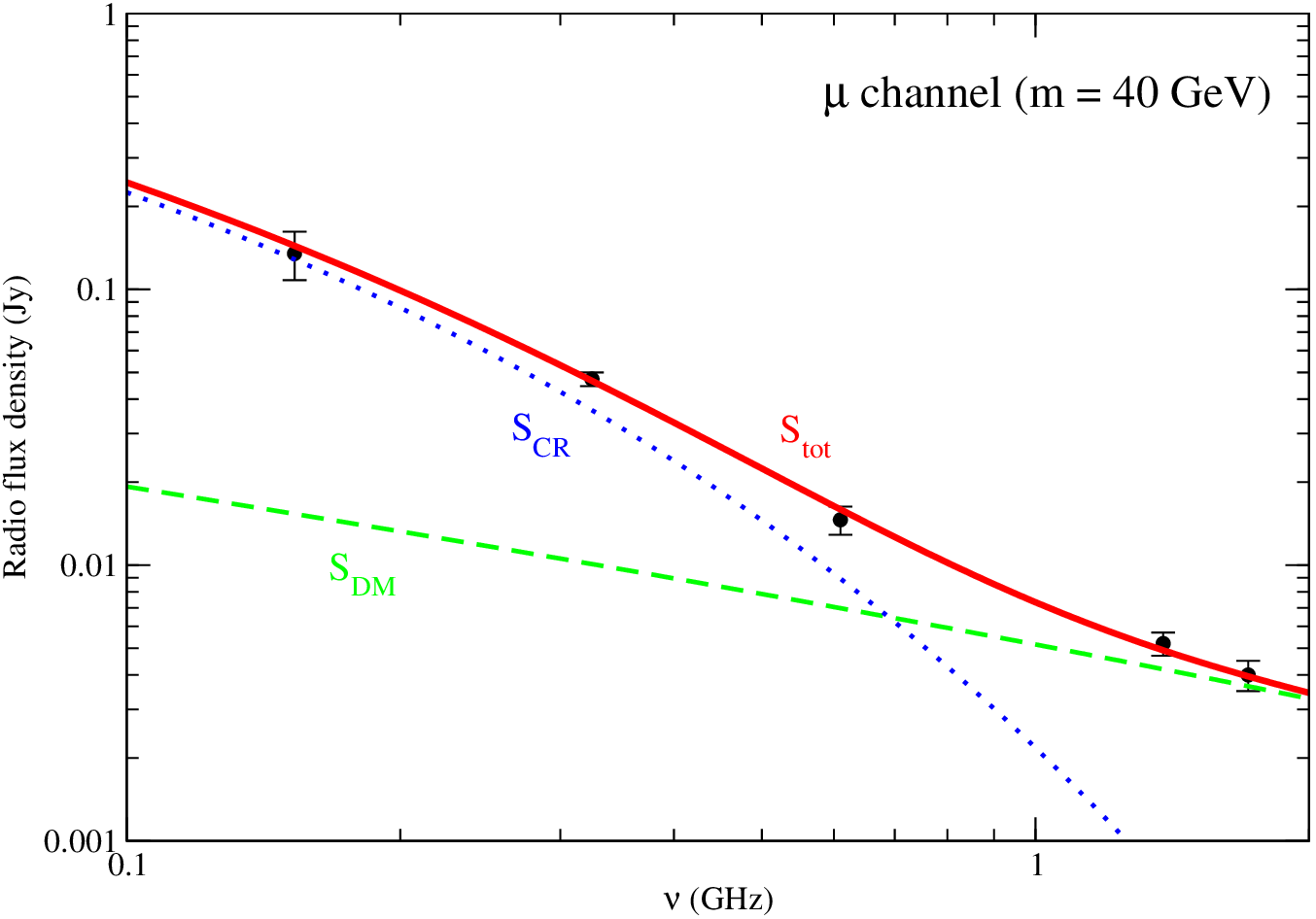} \newline
\newline
\newline
 \includegraphics[width=70mm]{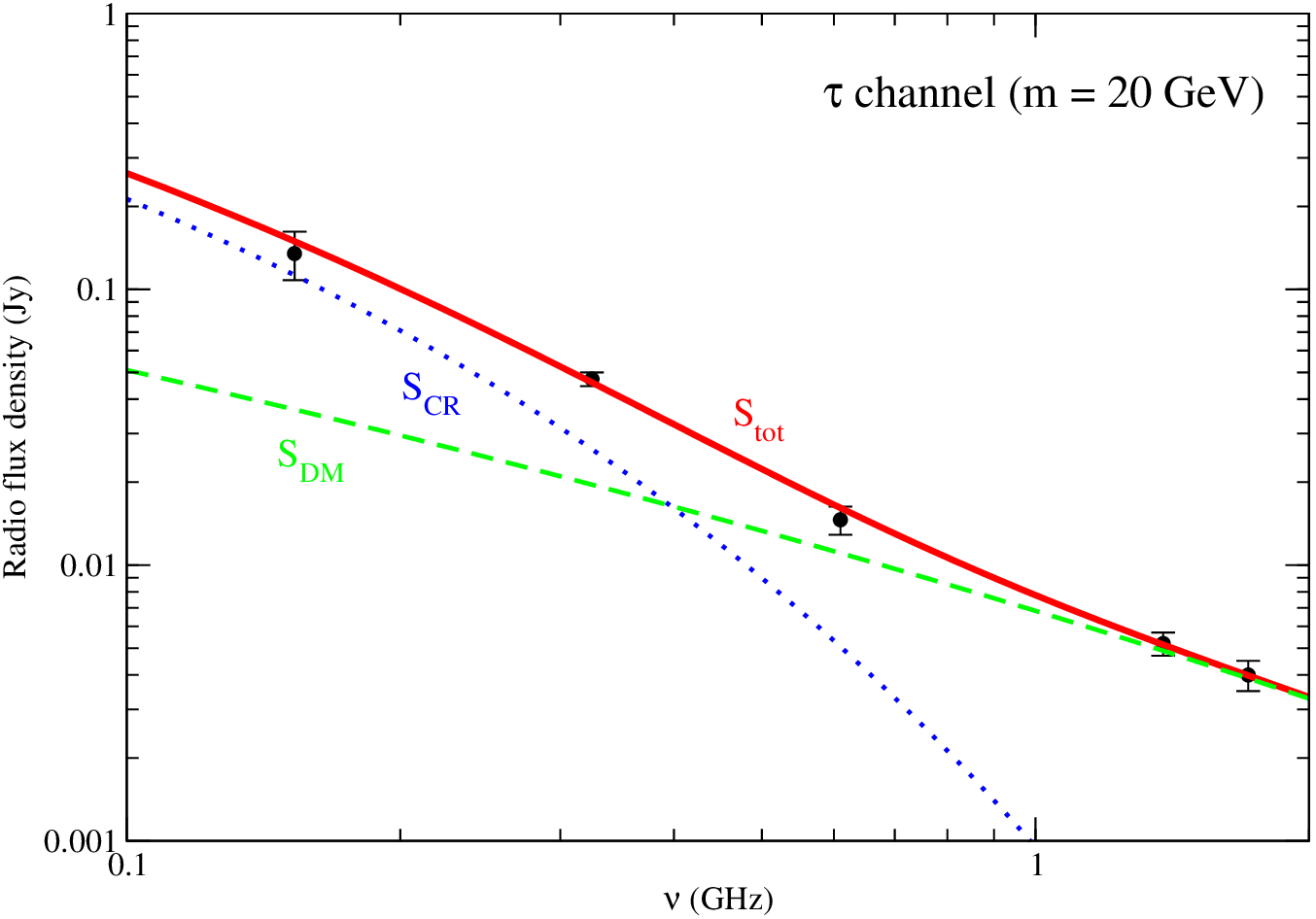}
~~ \includegraphics[width=70mm]{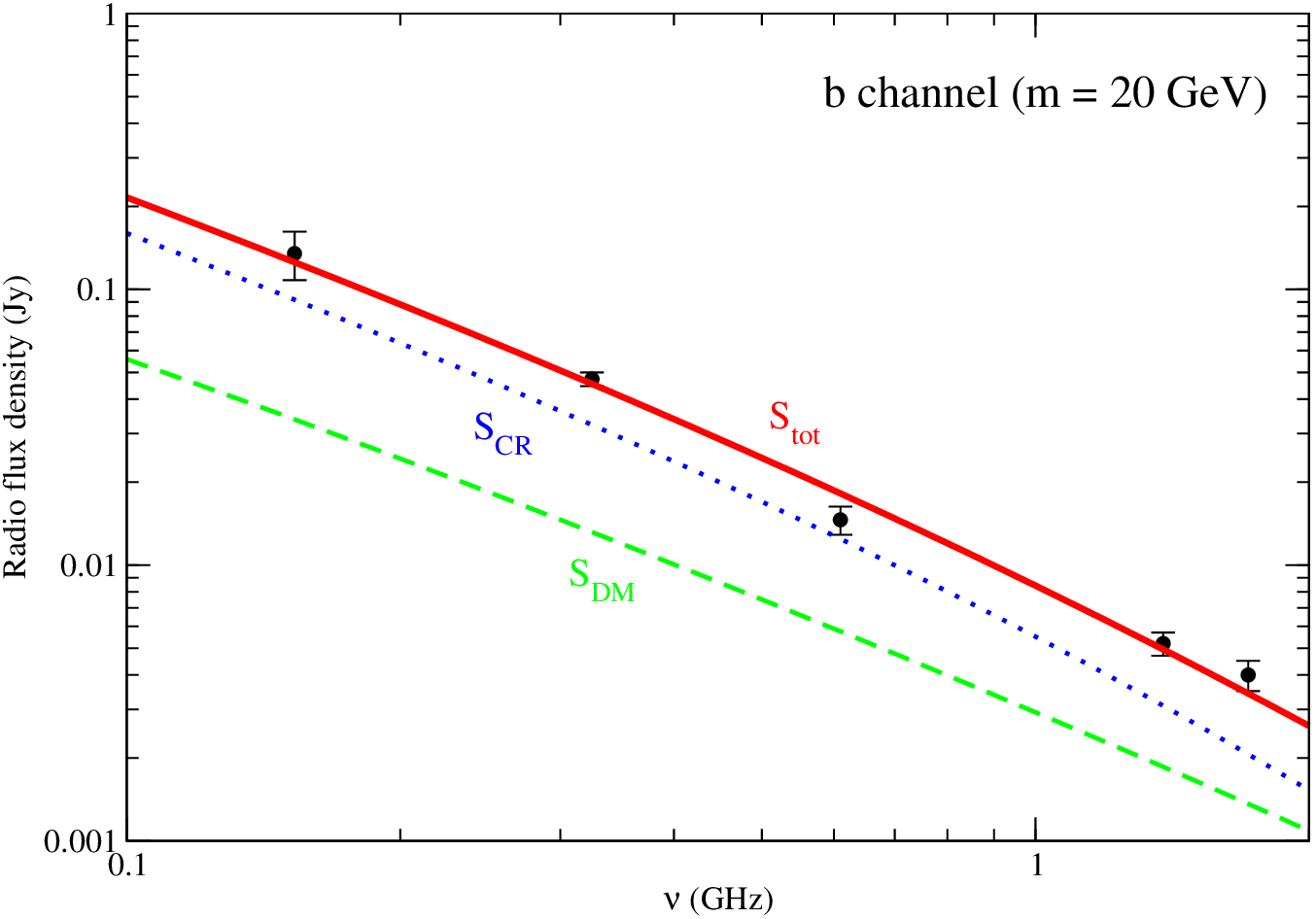} \newline
 \caption{The best-fit spectra for four annihilation channels (top left: $e^+e^-$ channel; top right: $\mu^+\mu^-$ channel; bottom left: $\tau^+\tau^-$ channel; bottom right: $b\bar{b}$ channel). The data with error bars are obtained from \citet{Macario}. The red solid lines are the total radio flux $S_{\rm tot}$. The blue dotted lines and green dashed lines are the contributions of cosmic rays $S_{\rm CR}$ (M3 model) and dark matter annihilation $S_{\rm DM}$ respectively. The corresponding best-fit parameters are shown in Table 1. Here, $B_0=15.7$ $\mu$G and $\eta=0.5$ are used.}
\vskip 10mm
\end{figure}

\begin{figure}
\vskip 10mm
 \includegraphics[width=70mm]{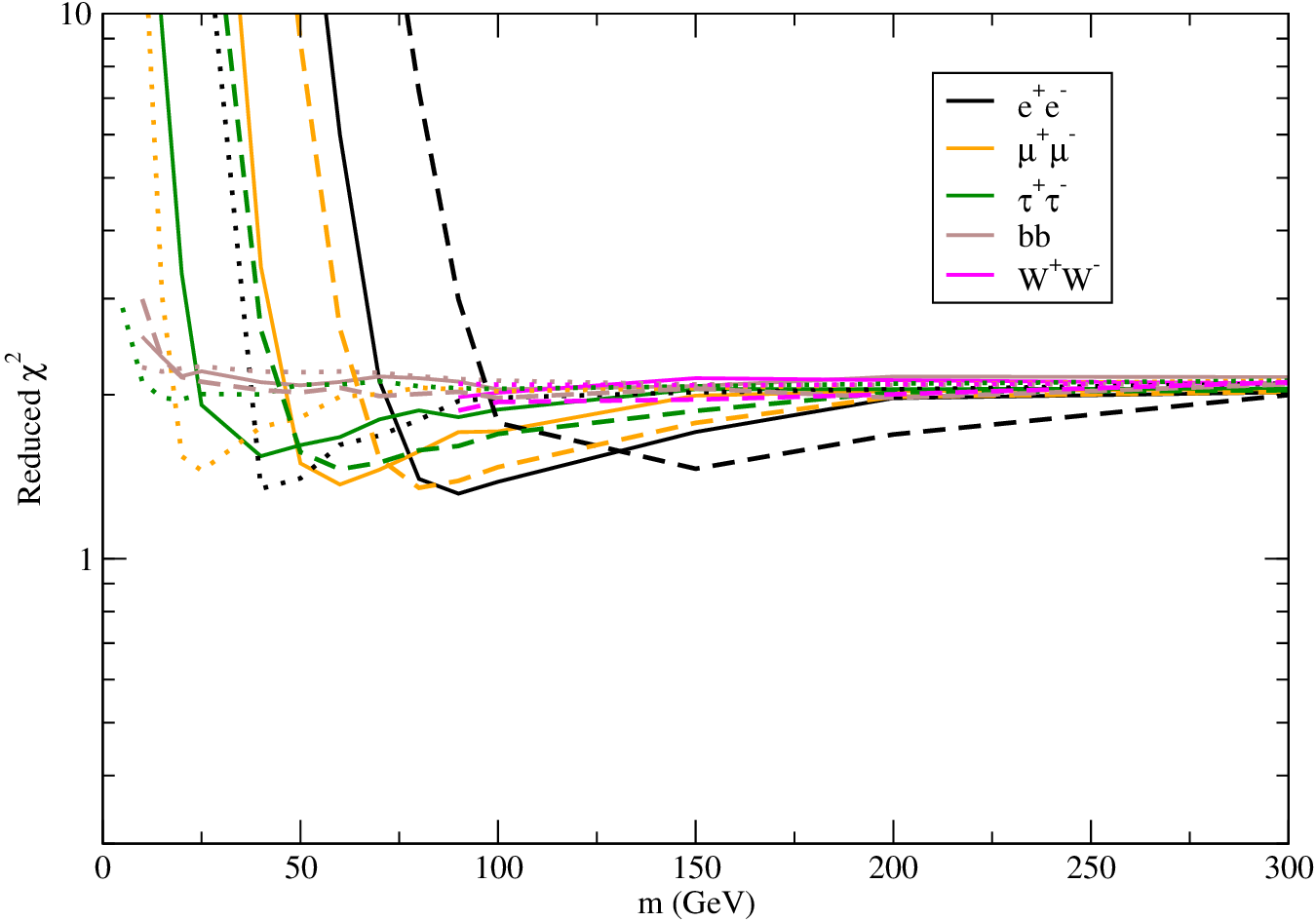}
~~ \includegraphics[width=70mm]{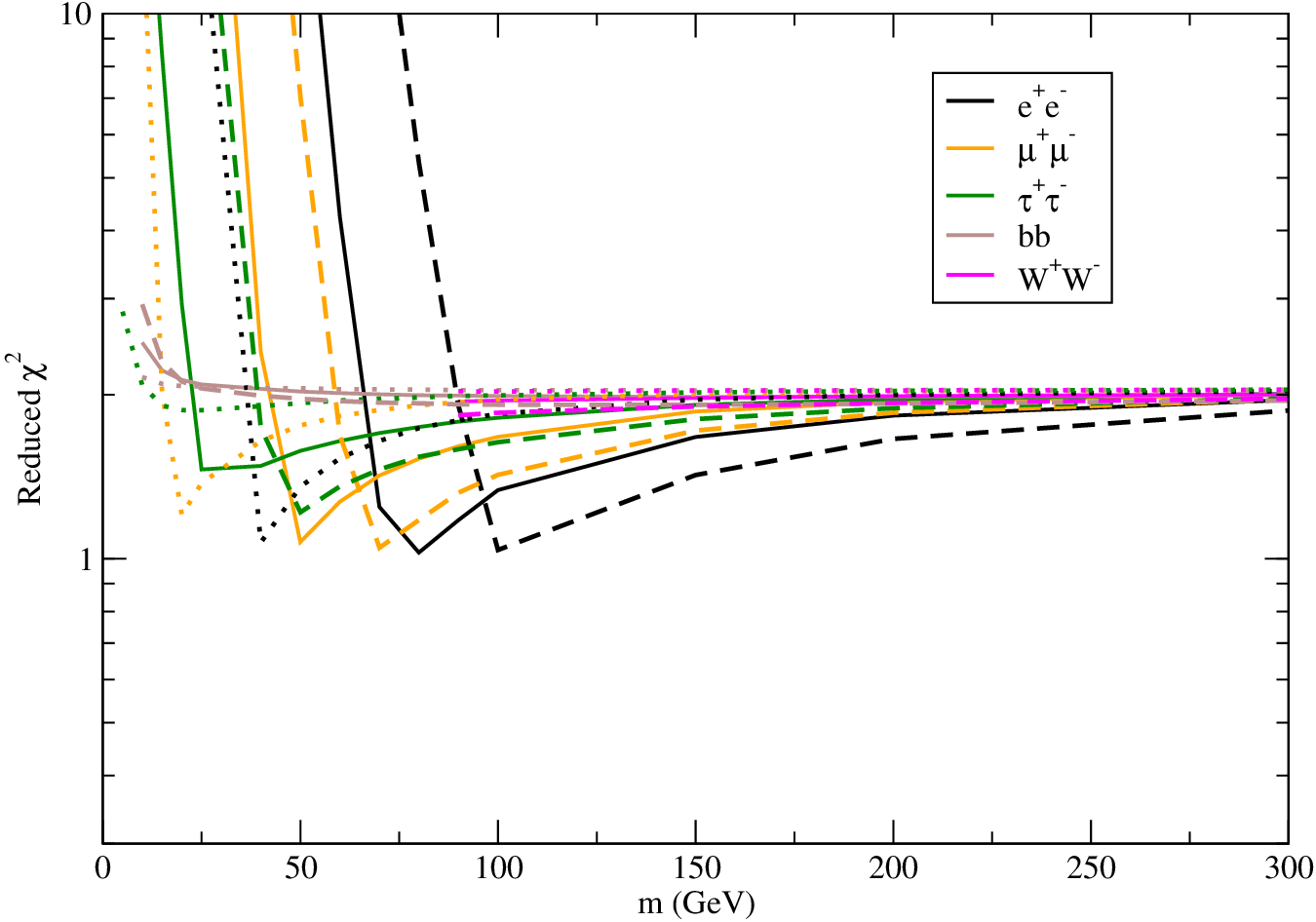} \newline
\newline
\newline
 \includegraphics[width=70mm]{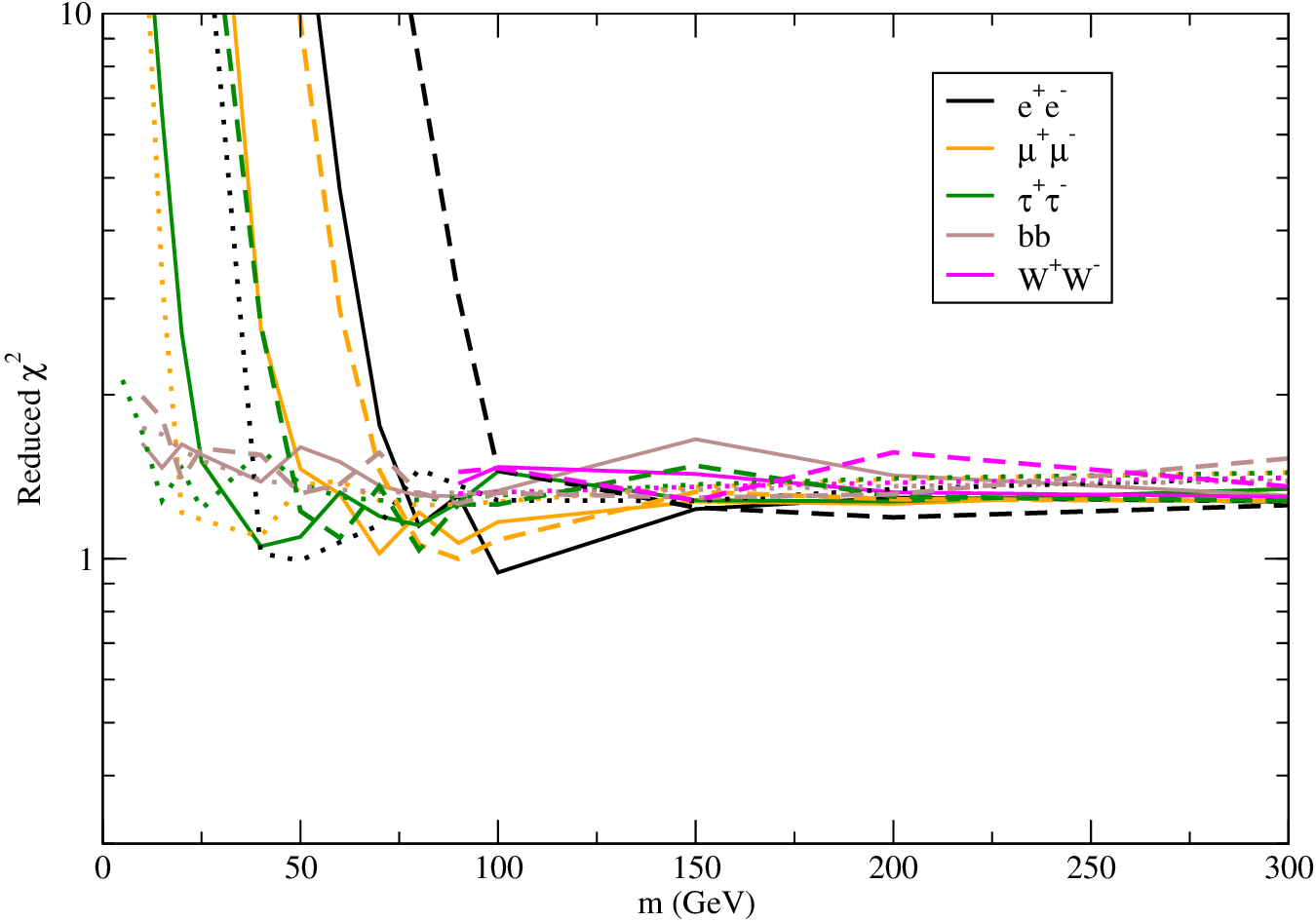}
~~ \includegraphics[width=70mm]{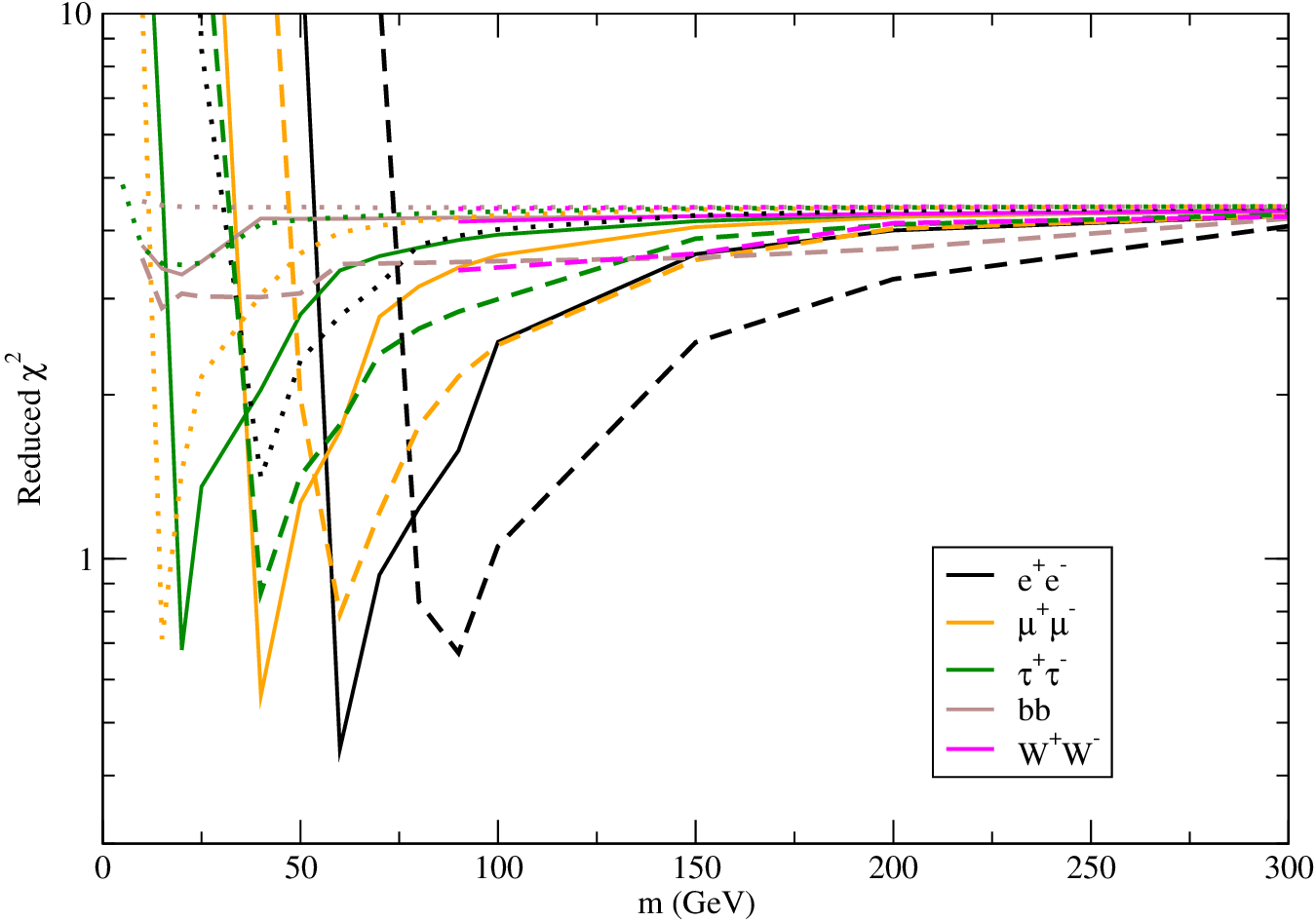}
 \caption{The minimum reduced $\chi^2$ values ($\chi_{\rm red}^2$) as a function of $m$ for five different annihilation channels (top left: M1 model with $\Gamma=0.5$; top right: M1 model with $\Gamma=1$; bottom left: M2 model; bottom right: M3 model). The dotted lines, solid lines, dashed lines represent the fits with the minimum dark matter density, average dark matter density and maximum dark matter density respectively. The range of the dark matter density (from minimum to maximum) is determined by the $1\sigma$ uncertainties of the hot gas parameters. Here, we take the hydrostatic density profile and assume $B_0=15.7$ $\mu$G and $\eta=0.5$.}
\vskip 10mm
\end{figure}

\begin{figure}
\vskip 10mm
 \includegraphics[width=70mm]{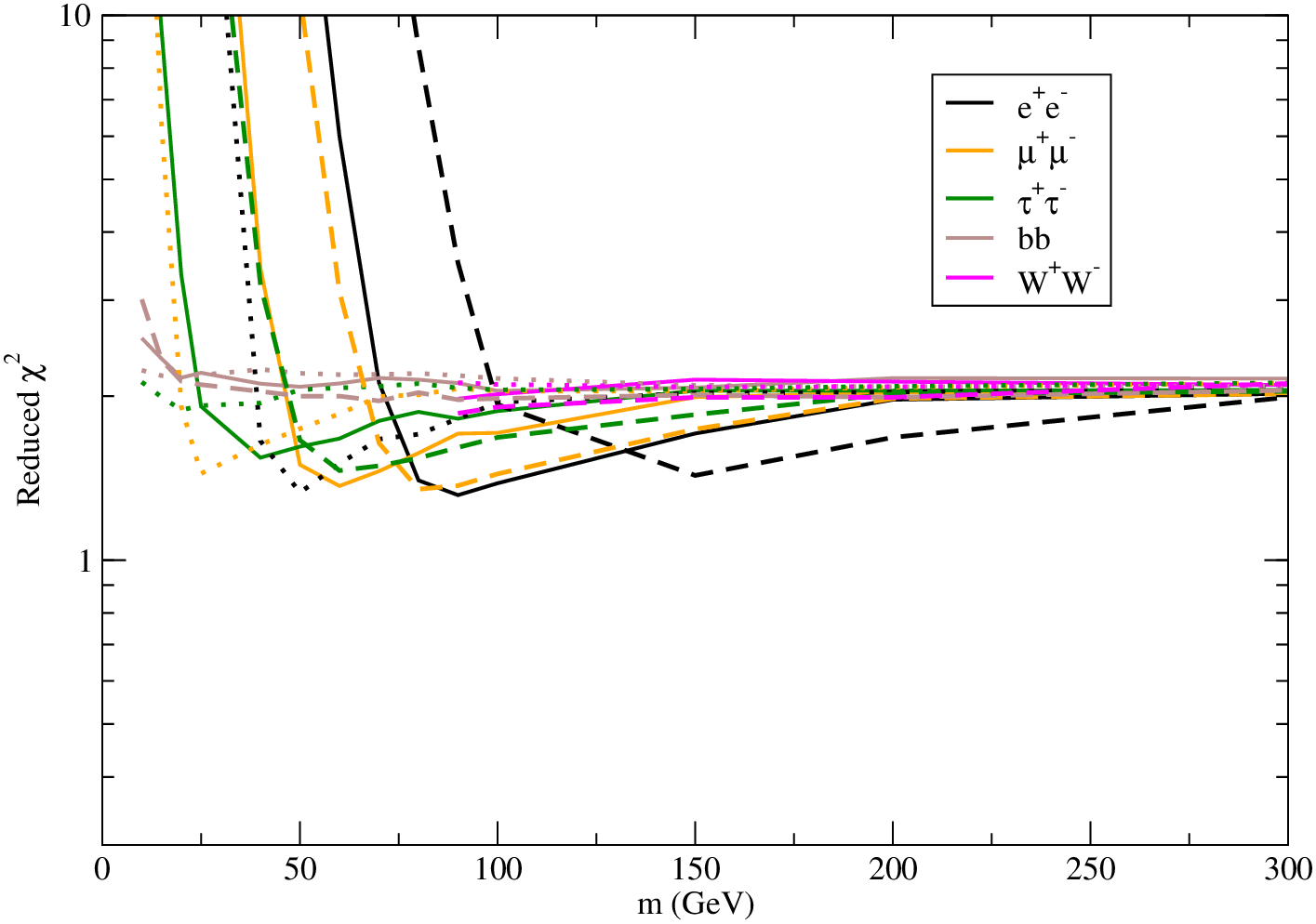}
~~ \includegraphics[width=70mm]{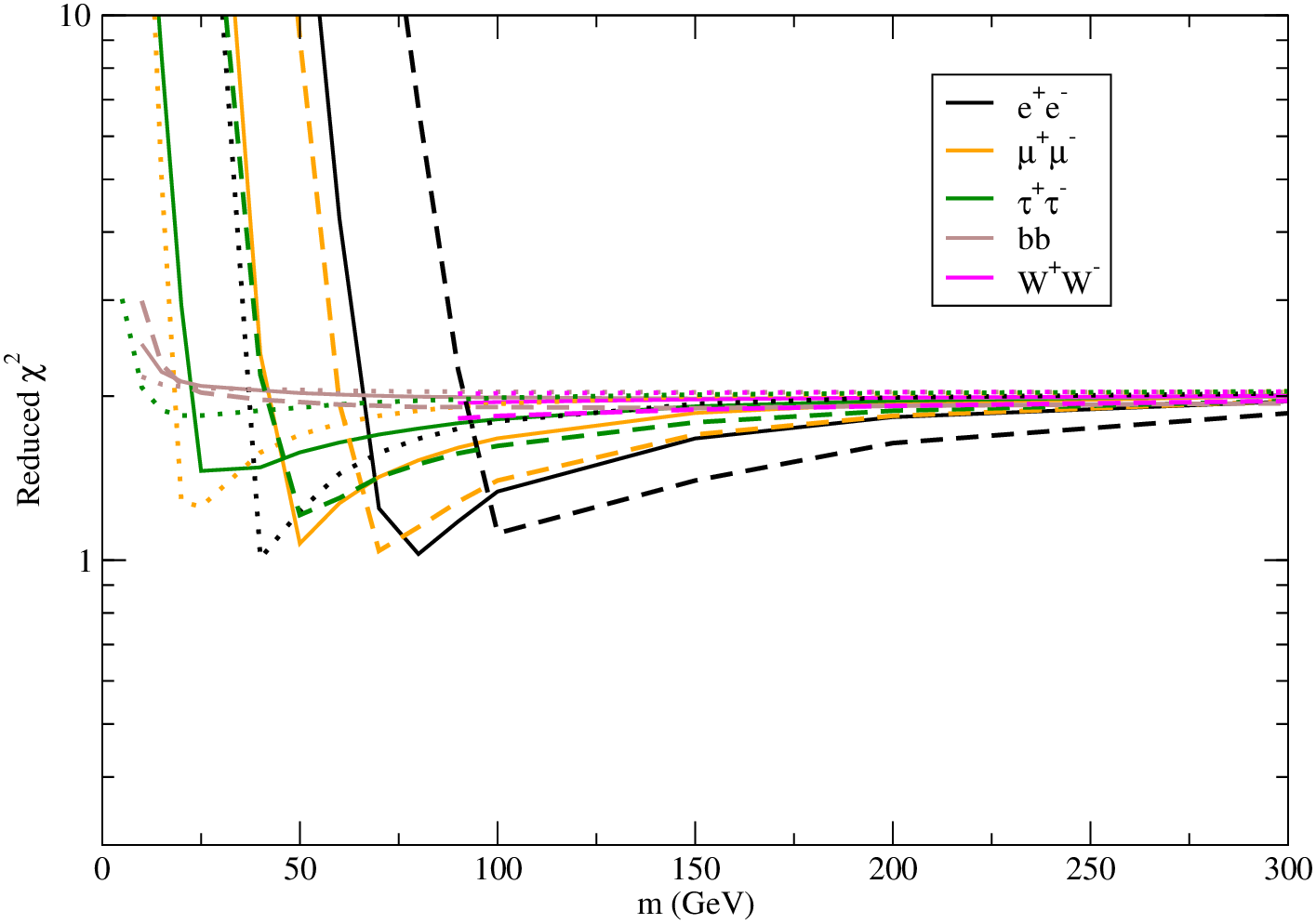} \newline
\newline
\newline
 \includegraphics[width=70mm]{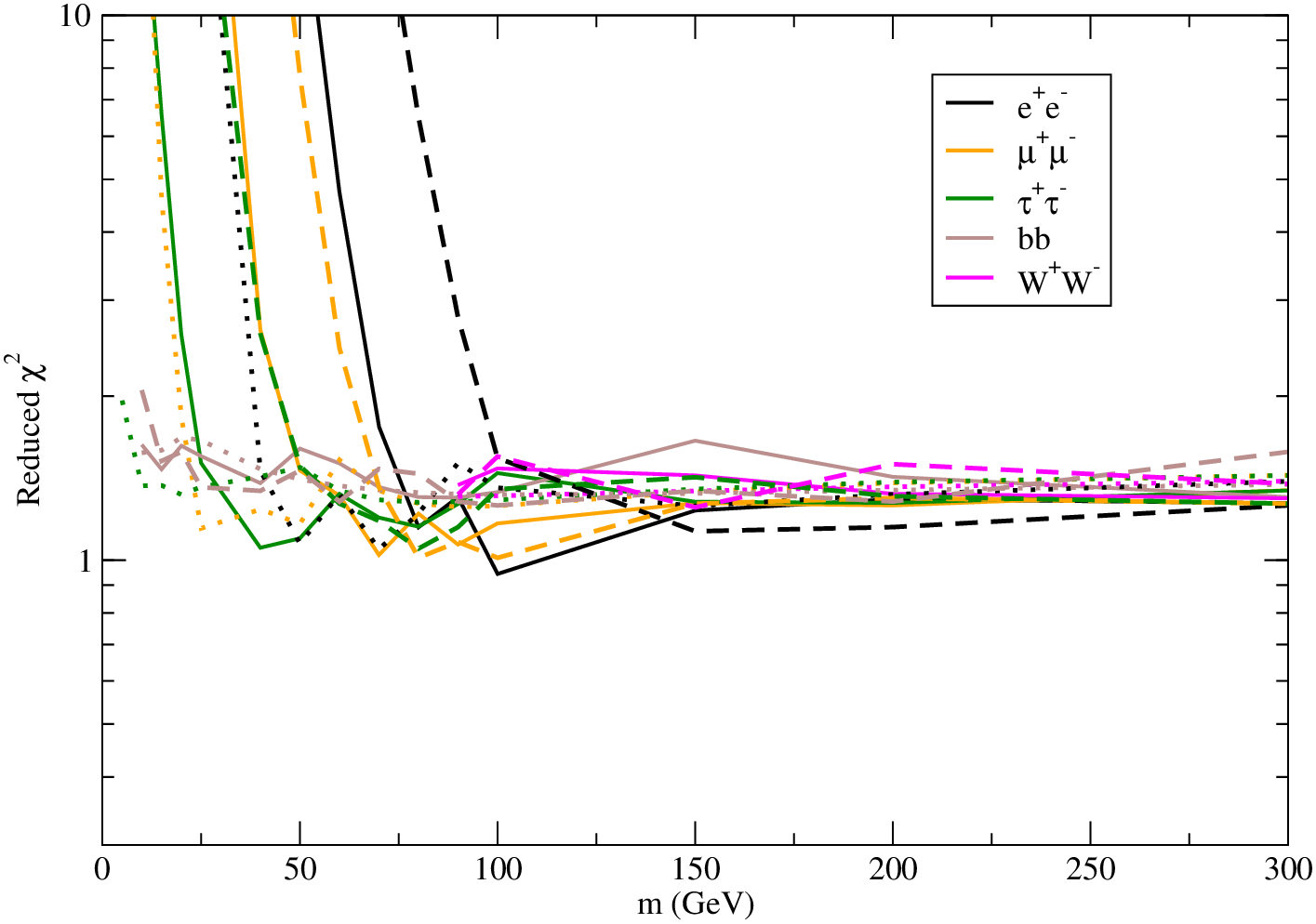}
~~ \includegraphics[width=70mm]{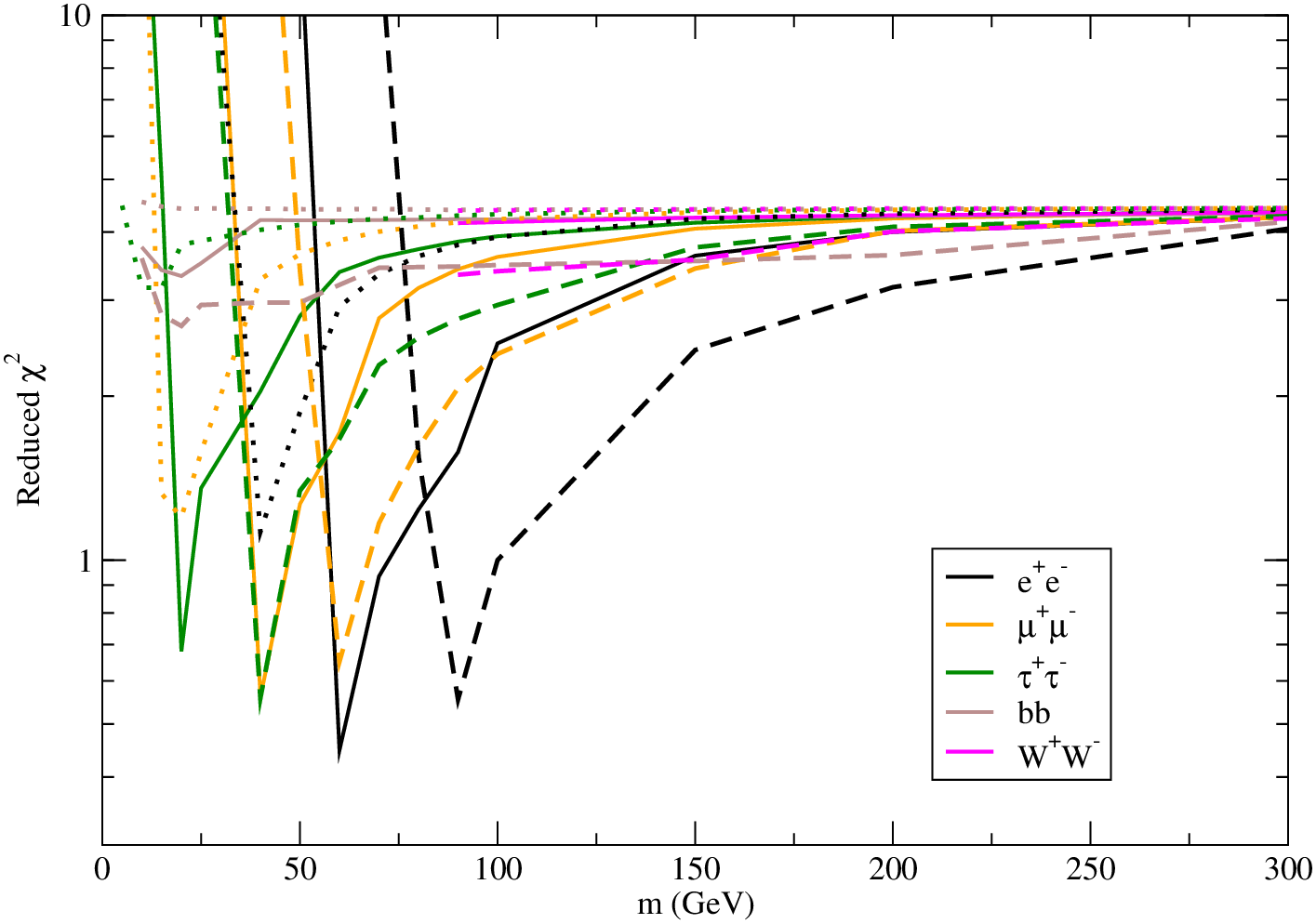}
 \caption{The minimum reduced $\chi^2$ values ($\chi_{\rm red}^2$) as a function of $m$ for five different annihilation channels (top left: M1 model with $\Gamma=0.5$; top right: M1 model with $\Gamma=1$; bottom left: M2 model; bottom right: M3 model). The dotted lines, solid lines, dashed lines represent the fits with $B_{\rm sub}=3.82$, $B_{\rm sub}=15.6$ and $B_{\rm sub}=31.2$ respectively. Here, we take the hydrostatic density profile and assume $B_0=15.7$ $\mu$G and $\eta=0.5$.}
\vskip 10mm
\end{figure}

\begin{figure}
\vskip 10mm
 \includegraphics[width=140mm]{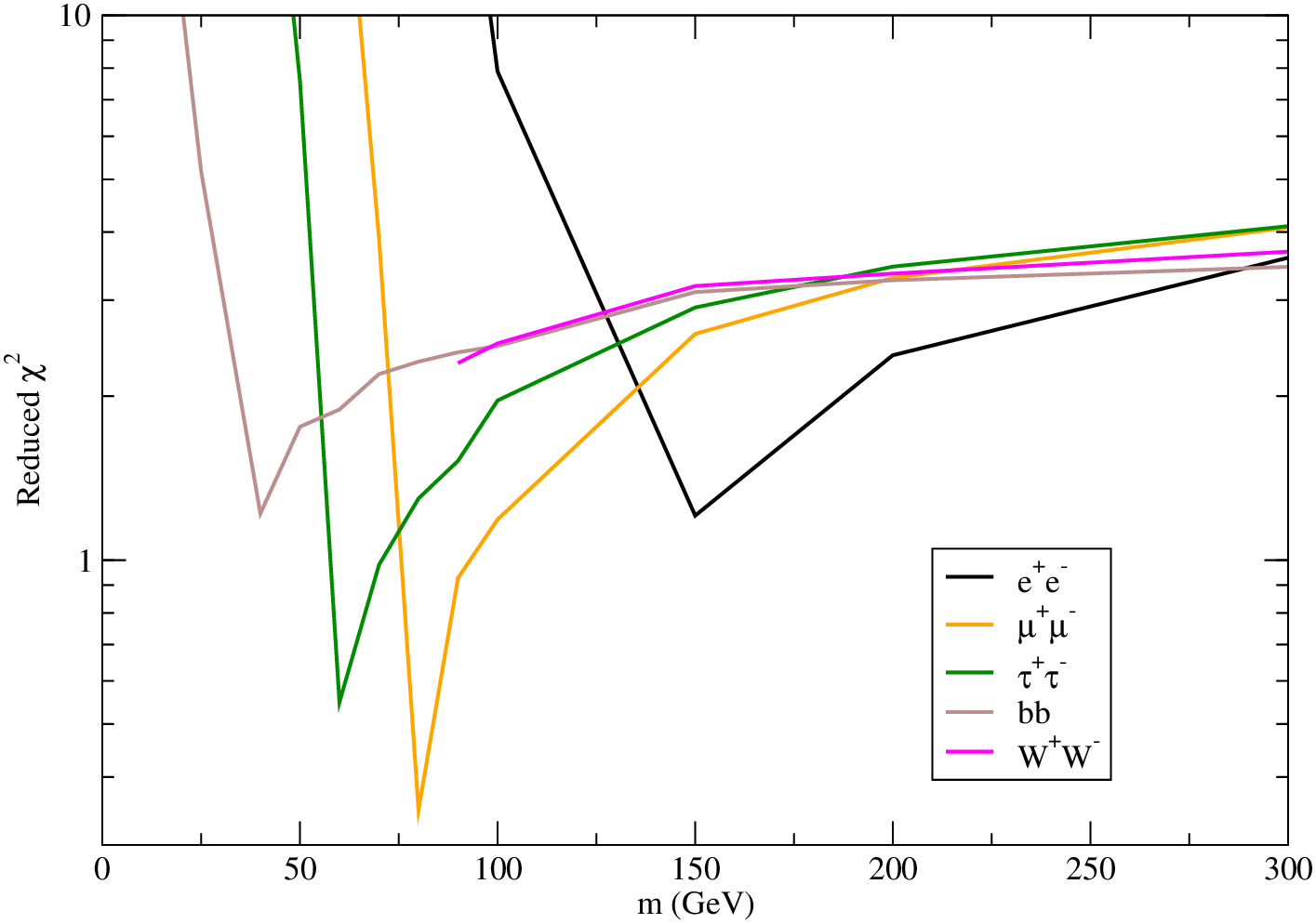}
 \caption{The minimum reduced $\chi^2$ values ($\chi_{\rm red}^2$) as a function of $m$ for five different annihilation channels (assuming the M3 model). We take the largest boost factor $B_{\rm sub}=31.2$ and the maximum limit of the dark matter density.}
\vskip 10mm
\end{figure}

\begin{table}
\caption{The best-fit parameters for four annihilation channels, assuming the hydrostatic density profile and $B_{\rm sub}=15.6$ neglecting the uncertainties of parameters.}
 \label{table1}
 \begin{tabular}{@{}lccccc}
  \hline
  Channel & $m$ (GeV) & $S_{\rm CR,0}$ (Jy) & $\alpha$ & $\nu_s$ (GHz) & $\chi_{\rm red}^2$ \\
  \hline
  $b\bar{b}$ & 20 & 0.02 & 1.08 & 0.60 & 3.3 \\
  $e^+e^-$ & 60 & 2.06 & 0.01 & 0.02 & 0.45 \\
  $\mu^+\mu^-$ & 40 & 0.70 & 0.30 & 0.03 & 0.56 \\
  $\tau^+\tau^-$ & 20 & 1.12 & 0.25 & 0.02 & 0.68 \\
  \hline
 \end{tabular}
\end{table}

\begin{acknowledgements}
The work described in this paper was supported by a grant from the Research Grants Council of the Hong Kong Special Administrative Region, China (Project No. EdUHK 28300518) and the Internal Research Fund from The Education University of Hong Kong (RG 2/2019-2020R).
\end{acknowledgements}

\end{document}